\documentclass[aps,pra,twocolumn,showpacs,superscriptaddress]{revtex4-1}  

\usepackage{graphicx}  
\usepackage{dcolumn}   
\usepackage{bm}        
\usepackage{amssymb}   
\usepackage{amsmath}

\usepackage[colorlinks=true,breaklinks=true,allcolors=blue]{hyperref}
\usepackage{url}
\usepackage{txfonts}

\usepackage{graphics}
\usepackage{pstricks}
\usepackage{tikz}
\usepackage{graphicx}
\usepackage{color}
\usepackage{hyperref}

\newcommand{\Eq}[1]{Eq.\,\eqref{eq:#1}}

\newcommand{\Fig}[1]{Fig.~\ref{fig:#1}}

\newcommand{\Sect}[1]{Sect.~\ref{sec:#1}}
\newcommand{\sect}[1]{\ref{sec:#1}}

\newcommand{\Tab}[1]{Table~\ref{tab:#1}}
\renewcommand{\vec}[1]{\mathbf{#1}}

\makeatletter
\let\cat@comma@active\@empty
\makeatother

\begin{document}

\hspace{5.2in} \mbox{Fermilab-Pub-04/xxx-E}

\title{Stability analysis of ground states in a one-dimensional trapped spin-1 Bose gas}
\author{C.-M. Schmied}
\affiliation{Kirchhoff-Institut f\"ur Physik,
            Ruprecht-Karls-Universit\"at Heidelberg, 
            Im~Neuenheimer~Feld~227,
            69120~Heidelberg, Germany}
\affiliation{Department of Mathematics and Statistics, 
                 University of Massachusetts, 
                 Amherst,
                 Massachusetts 01003-4515,
                 USA}
\author{T. Gasenzer}
\author{M. K. Oberthaler}
\affiliation{Kirchhoff-Institut f\"ur Physik,
            Ruprecht-Karls-Universit\"at Heidelberg, 
            Im~Neuenheimer~Feld~227,
            69120~Heidelberg, Germany}
\author{P. G. Kevrekidis}
\affiliation{Department of Mathematics and Statistics, 
                 University of Massachusetts, 
                 Amherst,
                 Massachusetts 01003-4515,
                 USA}
\date{\today}

\begin{abstract}
In this work we study the stability properties of the ground states of a spin-1 Bose gas in presence of a trapping potential in one spatial dimension. 
To set the stage we first map out the phase diagram for the trapped system by making use of a, so-called, continuous-time Nesterov method. 
We present an extension of the method, which has been previously
applied to one-component systems, to our multi-component system.  
We show that it is a powerful and robust tool for finding the ground states of a physical system without the need of an accurate  initial guess. 
We subsequently solve numerically the Bogoliubov de-Gennes equations
in order to analyze the stability of
the ground states of the trapped spin-1 system. 
We find that the trapping potential retains the overall structure of the
stability diagram, while affecting the spectral details of each of
the possible ground state waveforms.
It is also found that the peak density of the trapped system is the characteristic quantity describing dynamical instabilities in the system.
Therefore replacing the homogeneous density with the peak density of the trapped system leads to good agreement of the homogeneous Bogoliubov predictions with the numerically observed maximal growth rates of dynamically unstable modes.
The stability conclusions in the one-dimensional trapped system are independent of the spin coupling strength and the normalized trap strength over several orders of magnitude of their variation.
\end{abstract}

\pacs{%
}

\maketitle

\section{Introduction}

The study of coherent waveforms in the context of atomic physics has proved exceptionally productive over the past two decades,
especially so in the realm of Bose-Einstein condensates (BECs) \cite{Pitaevskii2003a, Kevrekidis2008a}.
Initially, the relevant effort focused on single-component condensates whereby in addition to the fundamental (ground) states, more exotic structures such as bright and dark solitons, as well as vortices and vortex rings/lines in higher dimensions were identified \cite{Kevrekidis2008a, Abdullaev2005, Frantzeskakis_2010, Bagnato2015}.
Very early on, a direction that emerged as being of interest in its own right due to its potential for phase transitions,
pattern formation and nonlinear wave structures was that of multi-component condensates. Binary mixtures
(typically of the same atomic species) in so-called pseudo-spinor BECs \cite{Hall1998a,Stamper-Kurn1998b} led to a remarkable array of developments, some of which have now been summarized in reviews in their own right \cite{KEVREKIDIS2016140}.

Aside from such pseudo-spinor BECs, a more recent development of even higher complexity has been the study of genuinely spinorial BECs~\cite{Stenger1999a}. 
In addition to more traditional topics, such as the ground states and their stability analysis (the so-called Bogoliubov-de Gennes
spectrum), the many body as well as excited states of such spinors have been identified \cite{Kawaguchi2012a.PhyRep.520.253}. 
Here, the topological excitations have been found to present unprecedented possibilities including not only fractional, but also non-Abelian vortices \cite{Kawaguchi2012a.PhyRep.520.253}.
In addition the study of spin-textures and transitions between them, phenomena including magnetic-dipole interactions and spin mixing have been of interest \cite{Stamper-Kurn2013a.RevModPhys.85.1191}. 
While these features have been summarized in some definitive reviews \cite{Kawaguchi2012a.PhyRep.520.253, Stamper-Kurn2013a.RevModPhys.85.1191}, spinor BECs remain a topic of active investigation including, e.g., the role of different types of coherent waveforms and solitonic structures, such as the recently experimentally identified dark-dark-bright and dark-bright-bright solitary waves \cite{PhysRevLett.120.063202}.

The ground states of this spinorial system have been mapped out in the absence of a trap \cite{Kawaguchi2012a.PhyRep.520.253}, and as a function of
some prototypical parameters such as the spinorial (i.e., spin-dependent
part of the) interaction strength or the so-called quadratic Zeeman energy shift (see below).
Nevertheless, far less is known about what happens in the presence, e.g., of an external trap, as is commonly the case in
experiments \cite{Pitaevskii2003a, Kevrekidis2008a}.
Moreover, while much effort has been devoted to numerical methods
aimed at identifying the ground (and excited) states of the single
component setting \cite{doi:10.1137/1.9780898719680}, this is far less so the case
in the context of multi-component systems.
It is the tackling of these types of questions that the present
study is devoted to.

We revisit the complex bifurcation diagram of the spin-1 Bose gas, proposing a very recently developed method (in the context
of one-component BECs in \cite{Ward2017a}). 
We find that the method performs {\it exceptionally well} towards the task of identifying ground states, even in the absence of a good initial guess for such a state. 
Moreover, we study these states in the presence of a parabolic
trap, varying not only the trap strength, but also the spin coupling
strength. 
We extend the homogeneous (no trap) case results to the
case where the trap is present, illustrating that the stability
properties of the one-dimensional system do not qualitatively change
and presenting a quantitative connection between the two. 
These results are useful as a benchmarking and extension of the standard
homogeneous case results \cite{Kawaguchi2012a.PhyRep.520.253}. 
At the same time, we feel that they offer useful tools for further computational exploration
of the system, including, e.g., excited states.

Our presentation is structured as follows. 
In \Sect{Model'}, we introduce the model, and its associated parameters. 
In \Sect{GroundStatesPhaseDiagram}, we discuss the ground states and phase diagram. 
In  \Sect{Stability}, we explore the stability analysis of the different states, while in
\Sect{Conclusion}, we summarize our results and present our conclusions including
the discussion of relevant directions for future study.

\section{The Model}
\label{sec:Model'}

We consider a one-dimensional (1D) spin-1 Bose-gas in a highly anisotropic trap with longitudinal and transverse trapping frequencies chosen such that $\omega_{\parallel} \ll \omega_{\perp}$. In that case the wave functions can be separated into a longitudinal and transverse part. The transverse wave function is the ground state of the respective harmonic oscillator and can be integrated out to obtain the following system of coupled 1D equations for the longitudinal part of the wave function

\begin{align}
\label{eq:EOM1}
i \hbar \partial_t \psi_{\pm 1} &= \mathcal{H}_0 \psi_{\pm 1} +  \tilde{q} \psi_{\pm 1} +  c_1^{(1\mathrm{D})} \left( \lvert \psi_{\pm 1} \lvert ^2  +\lvert \psi_{0} \lvert ^2 - \lvert \psi_{\mp 1} \lvert ^2\right) \psi_{\pm 1}  \nonumber \\  
& \quad + c_1^{(1\mathrm{D})} \psi_0^2 \psi_{\mp 1}^*  ,  
\end{align}

\begin{equation}
\label{eq:EOM2}
i \hbar \partial_t \psi_{0} = \mathcal{H}_0 \psi_{0} + c_1^{(1\mathrm{D})} \left( \lvert \psi_{1} \lvert ^2  + \lvert \psi_{- 1} \lvert ^2\right) \psi_{0} +2 c_1^{(1\mathrm{D})} \psi_{-1} \psi_{ 0}^* \psi_1. 
\end{equation}
Here, $\psi_{\pm 1}$ and $\psi_0$ are the bosonic fields which account for the magnetic sublevels $m_{\mathrm{F}} = \pm 1, 0$ of the $F = 1$ hyperfine manifold. 
The asterisk denotes the complex conjugate of the corresponding field. The spin-independent part of the Hamiltonian is given by $\mathcal{H}_0 = - [\hbar^2 / (2M) ] \partial_x^2 + \left(1/2\right) M \omega_{\parallel}^2 x^2 +c_0 \tilde{n}_{\mathrm{tot}}$, where $\tilde{n}_{\mathrm{tot}} = \lvert \psi_{1} \lvert ^2  +\lvert \psi_{0} \lvert ^2 + \lvert \psi_{-1} \lvert ^2$ is the total density and $M$ denotes the mass of the atoms.  
$\tilde{q}$ is the quadratic Zeeman energy shift which is proportional to an external magnetic field along the $z$-direction. It leads to an effective detuning of the $m_{\mathrm{F}} = \pm 1$ components with respect to the $m_{\mathrm{F}} = 0$ component. 
We are working in a frame where a possible homogeneous linear Zeeman shift has been absorbed into the definition of the fields. 
$c_0^{(1\mathrm{D})} = c_0/(2 \pi a_{\perp}^2)$ and $c_1^{(1\mathrm{D})} = c_1/(2 \pi a_{\perp}^2)$ with $a_{\perp} = \sqrt{\hbar / (M \omega_{\perp})}$ being the transverse harmonic oscillator length characterize the effectively one-dimensional density-density and spin-spin coupling.  
The coupling constants $c_0$ and $c_1$ are given by

\begin{equation}
c_0 = \frac{4 \pi \hbar^2 \left(a_0 + 2a_2 \right)}{3M}, \quad c_1 = \frac{4 \pi \hbar^2 \left(a_2 -a_0\right)}{3M},
\end{equation}
with the s-wave scattering lengths $a_0$ and $a_2$. For $c_1 < 0$ the interaction is ferromagnetic whereas for $c_1 > 0$ it is antiferromagnetic.

Measuring time, length and density in units of $\hbar/(c_0^{(1\mathrm{D})} n_{\mathrm{p}})$,
$[\hbar^2/ (M c_0^{(1\mathrm{D})} n_{\mathrm{p}})]^{1/2} $ and $n_{\mathrm{p}}$ respectively with $n_{\mathrm{p}}$ being the peak density of the system, we can write Eqs.~(\ref{eq:EOM1}) and (\ref{eq:EOM2}) in dimensionless form as

\begin{align}
\label{eq:EOM1Dimless}
i \partial_t \psi_{\pm 1} &= H_0 \psi_{\pm 1} +  q \psi_{\pm 1} +  \delta \left( \lvert \psi_{\pm 1} \lvert ^2  +\lvert \psi_{0} \lvert ^2 - \lvert \psi_{\mp 1} \lvert ^2\right) \psi_{\pm 1}  \nonumber \\  
& \quad + \delta \, \psi_0^2 \psi_{\mp 1}^*  ,  
\end{align}

\begin{equation}
\label{eq:EOM2Dimless}
i \partial_t \psi_{0} = H_0 \psi_{0} + \delta \left( \lvert \psi_{1} \lvert ^2  + \lvert \psi_{-1} \lvert ^2\right) \psi_{0} +2  \delta \,  \psi_{-1} \psi_{ 0}^* \psi_1, 
\end{equation}
where $H_0 = - \left(1/2 \right) \partial_x^2 + \left(1/2 \right) \Omega^2 x^2 + n_{\mathrm{tot}}$ with $n_{\mathrm{tot}} = \tilde{n}_{\mathrm{tot}}/n_{\mathrm{p}}$ and $q = \tilde{q} / (c_0^{(1\mathrm{D})} n_{\mathrm{p}})$. The normalized trap strength is 

 \begin{equation}
\label{eq:Omega}
\Omega = \frac {3} {2 \left( a_0 + 2 a_2\right) n_{\mathrm{p}}} \left (\frac{\omega_{\parallel}}{\omega_{\perp}} \right)
\end{equation}

and we define 

\begin{equation}
\delta = \frac {c_1^{(1\mathrm{D})}} {c_0^{(1\mathrm{D})}} = \frac{a_2 -a_0}{a_0 + 2a_2}.
\end{equation}
Typical values of $\delta \approx -5 \cdot 10^{-3}$ and $\delta \approx 3 \cdot 10^{-2}$ can be found in $^{87}$Rb and $^{23}$Na respectively; see,
e.g.,~\cite{PhysRevA.75.023617, PhysRevA.77.033612}.

\section{Ground states and phase diagram}
\label{sec:GroundStatesPhaseDiagram}

In this section we introduce the ground states and the corresponding mean-field phase diagram of the spin-1 Bose gas.
We start by discussing those properties in absence of a trapping potential.
To map out the phase diagram in presence of the trap we make use of a continuous-time Nesterov (CTN) scheme.
We present an extension of the scheme, previously applied to one-component systems, to our multi-component system in the third part of the section.
In the final part of this section we show numerical results obtained with this method for the one-dimensional
trapped spin-1 Bose gas. 

\subsection{Mean-field phase diagram for a homogeneous system}
\label{sec:MeanFieldPhaseDiag}

\begin{figure}
\includegraphics{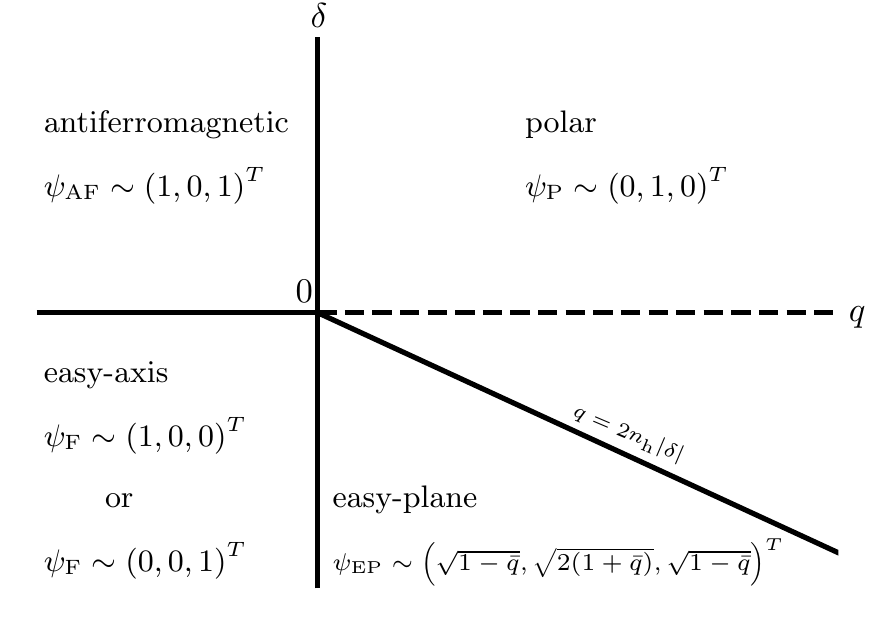}
\caption{\label{fig:PhaseDiag} Mean-field phase diagram of the spin-1 Bose gas in absence of a trapping potential ($\Omega = 0$) for vanishing z-component of the magnetization (we are implicitly accounting here for the fact that the ground state in the easy-axis phase is degenerate). 
For $\delta > 0$ we obtain two different phases. 
For $q>0$ the system is in the polar phase whereas for $q<0$ the system is in the antiferromagnetic phase. 
The phase transition occurs at $q =0$. 
For $\delta < 0$ three phases exist. In case of $ q > 2 n_{\mathrm{h}} \lvert \delta \rvert$ the system is in the polar phase, for $0 < q < 2 n_{\mathrm{h}} \lvert \delta \rvert$ it is in the easy-plane phase and for $q<0$ it is in the easy-axis phase. 
A quantum phase transition (QPT) occurs at $q = 2 n_{\mathrm{h}} \lvert \delta \rvert$. 
The phase transition between the easy-plane and easy-axis phase is at $q =0$. 
All phase transitions are marked by black solid lines. 
Note that $\bar{q} = q/ (2 n_{\mathrm{h}} \lvert \delta \rvert)$ and $n_{\mathrm{h}}$ is the homogeneous total density. 
For details on the ground states see main text.}
\end{figure}

In the following we give a brief review of the mean-field phase diagram of the spin-1 Bose gas in a homogeneous system ($\Omega =0$) which has been discussed
in detail in \cite{Kawaguchi2012a.PhyRep.520.253}. 
A schematic plot of the phase diagram is shown in Fig.~\ref{fig:PhaseDiag}.

(i) \emph{polar phase} --
For $\delta >0$ the  spin interaction is antiferromagnetic. 
If $q>0$ additionally, the system is in the polar phase. 
The ground state is unmagnetized and it is given by the state vector 

\begin{equation}
\label{eq:PolarState}
\vec{\psi}_{\mathrm{P}} = e^{i \theta} \sqrt{n_{\mathrm{h}}}
\begin{pmatrix}
0 \\ 1 \\ 0
\end{pmatrix}.
\end{equation}
Here, $n_{\mathrm{h}}$ is the homogeneous total density of the system and $\theta$ is a global phase distinguishing different realizations of the spontaneous symmetry breaking. 

(ii) \emph{antiferromagnetic phase} --
If $q<0$, but $ \delta >0$,  the system is in the antiferromagnetic phase. 
The ground state is again unmagnetized and its state vector reads

\begin{equation}
\vec{\psi}_{\mathrm{AF}} = \sqrt{n_{\mathrm{h}}}
\begin{pmatrix}
 e^{i \theta_1} \\ 0 \\  e^{i \theta_{-1}}
\end{pmatrix}.
\end{equation}
Here, $\theta_{\pm1}$ are arbitrary phases of the $m_{\mathrm{F}} = \pm 1$ components.
The first-order phase transition separating the antiferromagnetic and polar phase occurs at $q =0$. 

(iii) \emph{easy-axis phase} --
For $\delta <0$, 
the  spin interaction is ferromagnetic.
If $q < 0$ additionally, the system is in the easy-axis ferromagnetic phase. 
The two degenerate ground states emerge by an explicit symmetry breaking in the $m_{\mathrm{F}} = \pm 1$ components. 
This leads to a state which is either fully magnetized in $+z$ or $-z$ direction, i.e. $f_z = \left (n_1 -n_{-1} \right) / n_{\mathrm{h}} = \pm1$. The corresponding state vectors are given by

\begin{equation}
\vec{\psi}_{\mathrm{F}} = e^{i \theta} \sqrt{n_{\mathrm{h}}}
\begin{pmatrix}
1 \\ 0 \\ 0
\end{pmatrix} \quad   \mathrm{or} \quad 
\vec{\psi}_{\mathrm{F}} = e^{i \theta} \sqrt{n_{\mathrm{h}}}
\begin{pmatrix}
0 \\ 0 \\ 1
\end{pmatrix}.
 \end{equation}
At $q=0$ a first-order phase transition occurs in the system.

(iv) \emph{easy-plane phase} --
For $\delta < 0$ and $0 < q <  q_0$ with $q_0 = 2 n_{\mathrm{h}} \lvert \delta \rvert$ 
the system is in the easy-plane ferromagnetic phase in which the mean-field ground state reads
\begin{equation}
  \vec{\psi}_{\mathrm{EP}} =  \sqrt{n_{\mathrm{h}}} 
  \frac{e^{i \theta}} 2 
  \begin{pmatrix}
  e^{-i \phi}\sqrt{ 1 - q/q_0}   \\ \sqrt{  2(1 + q/q_0) }  \\ e^{i \phi} \sqrt{ 1 - q/q_0}  
  \end{pmatrix},
\end{equation}
where $\phi$ denotes the angle with respect to the spin-$x$-axis. 
The complex order parameter in the easy-plane phase is the transversal spin $F_{\perp} = F_x + i F_y = \sqrt{2} \left( \psi_0^* \psi_1 + \psi_{-1}^* \psi_0 \right) $.
The ground state gives rise to the mean spin vector lying in the transversal spin plane, with magnetization $ \lvert f_{\perp}\rvert = \lvert F_{\perp} \rvert/ n_{\mathrm{h}} = [1- (q/q_0)^{2} ]^{1/2}$. 
At $q = q_0$ the system exhibits a quantum phase transition (QPT) that breaks the full spin symmetry of the ground state. 
For $q>q_0$ the system is again in the polar phase with the unmagnetized ground state given by Eq.~(\ref{eq:PolarState}).

\subsection{Time-independent equations of motion}
\label{sec:TIEOM}

In the remainder of this section we want to determine the ground states of the system in presence of a trapping potential, i.e., we aim to identify the stationary states of the trapped system with the lowest eigenenergy.
By choosing the general ansatz $\psi_m (x,t) = \psi_m (x) e^{-i \mu_m t}$ with $m = 0, \pm 1$ and $\mu_m$ being the chemical potential of each spinor component, Eqs.~(\ref{eq:EOM1Dimless}) and (\ref{eq:EOM2Dimless}) turn into

\begin{align}
\label{eq:TimeDepEomWithMuPM1}
\mu_{\pm 1} \psi_{\pm 1} &= H_0 \psi_{\pm 1} +  q \psi_{\pm 1} +  \delta \left( \lvert \psi_{\pm 1} \lvert ^2  +\lvert \psi_{0} \lvert ^2 - \lvert \psi_{\mp 1} \lvert ^2\right) \psi_{\pm 1}  \nonumber \\  
& \quad + \delta \, \psi_0^2 \psi_{\mp 1}^*  e^{-i (2 \mu_0 - \mu_1 -\mu_{-1}) t} ,  
\end{align}

\begin{align}
\label{eq:TimeDepEomWithMu0}
\mu_0 \psi_{0} &= H_0 \psi_{0} + \delta \left( \lvert \psi_{1} \lvert ^2  + \lvert \psi_{-1} \lvert ^2\right) \psi_{0} \nonumber \\
& \quad + 2  \delta \, \psi_{-1} \psi_{ 0}^* \psi_1  e^{-i (\mu_1 + \mu_{-1} - 2\mu_0) t}. 
\end{align}
A stationary state resulting from Eqs.~(\ref{eq:TimeDepEomWithMuPM1}) and (\ref{eq:TimeDepEomWithMu0}) has to fulfill the phase matching condition $2 \mu_0 - \mu_1 - \mu_{-1} = 0$.   
As a population imbalance between the $m_{\mathrm{F}} = \pm 1$ components is not favored, independent of the choice of the couplings in the equations of motion, we assume that $\mu_1 = \mu_{-1}$ for all stationary states considered in this work. 
This implies that $\mu_0 = \mu_1 = \mu_{-1}  \equiv \mu$. 
The time-independent equations of motion thus read:

\begin{align}
\label{eq:EOM1TI}
\mathcal{F}_{\pm 1} (\psi_1, \psi_0, \psi_{-1}, \psi_1^*,  \psi_0^*,  \psi_{-1}^*) \equiv &- \mu \psi_{\pm 1} + H_0 \psi_{\pm 1} +  q \psi_{\pm 1}   \nonumber \\ & +  \delta \left( \lvert \psi_{\pm 1} \lvert ^2  +\lvert \psi_{0} \lvert ^2 - \lvert \psi_{\mp 1} \lvert ^2\right) \psi_{\pm 1}  \nonumber \\ &+   \delta \, \psi_0^2 \psi_{\mp 1}^* \nonumber \\ =& \, 0 ,  
\end{align}

\begin{align}
\label{eq:EOM2TI}
\mathcal{F}_0  (\psi_1, \psi_0, \psi_{-1}, \psi_1^*,  \psi_0^*,  \psi_{-1}^*) \equiv& -\mu \psi_{0} + H_0 \psi_{0}
\nonumber \\&+ \delta \left( \lvert \psi_{1} \lvert ^2  + \lvert \psi_{-1} \lvert ^2\right) \psi_{0}  \nonumber \\ &+2  \delta \, \psi_{-1} \psi_{ 0}^* \psi_1 \nonumber \\ =& \, 0. 
\end{align}
Here, we introduced functions $\mathcal{F}_{0, \pm 1}$ as abbreviations for the time-independent equations of motion which will be of practical use in \Sect{BdG}. 

Various first- and second-order methods can be applied to find solutions to the above stated equations of motion. 
A commonly used method for such a problem is an exact Newton scheme.
It is a second-order method  which involves the explicit calculation of the Jacobian.
A major advantage of the Newton scheme is that it is not restricted to finding ground states (i.e.~the global energy minimum) of a physical system. 
On the other hand, a disadvantage of this scheme is that
an adequate(ly proximal to the true solution) initial guess for the
wave functions is needed to ensure convergence.
For the trapped spin-1 system, however, a priori, we do not know where the quantum phase transition (QPT) between the easy-plane and polar phase is located. 
Moreover, our initial guesses in the trapped case may not be sufficiently accurate. 
Thus, the Newton method might fail to converge to the true ground state.
In that light, in what follows, we first focus on a method which is able to map out the spin-1
phase diagram {\it independently} of the
specifics of the initial guess.
Nonetheless, we will make use of the Newton method later on in order to find a specific state of interest also within a phase where it is not 
the ground state anymore. 
This is required to perform the stability analysis for a given state throughout the whole $(\delta,q)$-plane of the spin-1 
phase diagram.
The Newton scheme for the spin-1 system will be discussed in detail in \Sect{Newton}.

\subsection{Continuous-time Nesterov scheme}
\label{sec:CTN}

For simplicity we assume that we are interested in the variational problem of minimizing the function $G(x)$. 
Following the classical discrete-time Nesterov (mirror descent) algorithm~\cite{Nesterov1983}, it has been shown
in the work of~\cite{JMLR:v17:15-084} that one can formulate a continuous-time analogue. 
This involves a second-order ordinary differential equation (ODE), which in some sense generalizes
standard gradient descent schemes.  
The ODE is given by

\begin{equation}
\label{eq:ODE}
\ddot{x} + \frac {3}{t}  \dot{x} + \frac{d}{dx} G(x) = 0,
\end{equation}
where the dots denote derivatives with respect to the continuous time
variable $t$. 
We refer to this scheme as the continuous-time Nesterov (CTN) method.
In general, Eq.~(\ref{eq:ODE}) can be viewed as describing the damped motion of a particle in a potential $G(x)$.
In contrast to standard gradient descent schemes we are dealing with a second-order differential equation resulting in the ``acceleration
vector'' pointing into the direction of the steepest descent.
The strength of the damping $\sim t^{-1}$ explicitly depends on the evolution time, i.e.~the damping is large at small times when the particle
is, comparatively, further away from the fixed point solution and decreases as the fixed point solution is approached, which is ensured by choosing an appropriate time step as well as a proper preconditioner when solving Eq.~(\ref{eq:ODE}) numerically.
The preconditioner is an operator (or upon discretization, a matrix) that helps solving the linear system at hand by reducing its condition number.

The CTN method can be optimized by introducing
a so-called gradient restarting scheme \cite{JMLR:v17:15-084,Ward2017a}.
Following that scheme in two or three spatial dimensions, the time $t$ is reset to 1 when the angle between
the negative gradient $- \nabla G(\vec{x})$ of the function $G$ that
we are trying to extremize and $\dot{\vec{x}}$ is larger than 90 degrees and a pre-specified amount of time $t_{\mathrm{res}}$ has elapsed. 
In one spatial dimension this geometrical condition boils down to the inner product of $- dG(x)/dx$ and $\dot{x}$ being smaller than 0.
Gradient restarting ensures that the CTN is sufficiently damped in all stages of the evolution. 
This can be intuitively understood when thinking again of the motion of a particle in a potential. 
The inner product of $- dG/dx$ and $\dot{x}$ being smaller than 0 means that our particle is moving in the direction of the fixed point solution. 
To avoid possible oscillations of the solution in the vicinity
of the fixed point in case of weak damping we then reset the time which results in a large damping of the motion.
Note that gradient restarting is only useful when the specific geometrical condition stated above is fulfilled.
We refer to the optimized scheme as accelerated continuous-time Nesterov (ACTN) method.

In Ref.~\cite{Ward2017a}, it has recently been shown that the CTN method and its accelerated version can also be applied to functionals and can be used for  finding stationary states of partial differential equations (PDEs). In particular, the CTN method
was utilized for analyzing stationary states in a one-component Bose gas in
one and two spatial dimensions. 
Here, we extend  the CTN  method to a multi-component system, illustrating
its ability to capture ground states in a wide range of parametric
regimes and rather independently of the specifics of the initial
guess. Following the steps of \cite{Ward2017a},
which are based on the replacements $x \rightarrow \psi (x)$ and $dG/dx \rightarrow \mathcal{F}(\psi(x))$, i.e., replacing derivatives with respect to the spatial coordinate $x$ by functional derivatives with respect to the field $\psi (x)$, we obtain a PDE for the evolution (towards equilibrium) of the field $\psi$ in space and time.
Generalizing the replacements for all hyperfine components, the CTN scheme for the spin-1 system assumes the form:

\begin{align}
0 =&\,\,  \ddot{\psi}_{\pm 1} + \frac{3}{t} \dot{\psi}_{\pm 1}  -  \left[ - H_0 \psi_{\pm 1}  -  q \psi_{\pm 1}  \right. \nonumber \\
 &\left. - \,\delta \left( \lvert \psi_{\pm 1} \lvert ^2   + \lvert \psi_{0} \lvert ^2 - \lvert \psi_{\mp 1} \lvert ^2\right) \psi_{\pm 1}  - \, \delta \, \psi_0^2 \psi_{\mp 1}^* + \mu \psi_{\pm 1}  \right],  
\end{align}

\begin{align}
 0 =& \,\, \ddot{\psi}_{0} + \frac{3}{t} \dot{\psi}_{0} - \left [ -H_0 \psi_{0} - \delta \left( \lvert \psi_{1} \lvert ^2  + \lvert \psi_{-1} \lvert ^2\right) \psi_{0} \right.\nonumber \\ &\left.- \, 2  \delta\,  \psi_{-1} \psi_{ 0}^* \psi_1 + \mu \psi_{0} \right].
\end{align}

Note that the overdot denotes a partial derivative with respect to time, as we are performing
a distributed minimization by solving the partial differential equations
of the above system.
We use a second-order center difference scheme for approximating the second derivative and a first-order backward difference scheme for approximating the first derivative with respect to time.
This leads to the following  evolution equations 

 \begin{align}
\psi_{\pm 1}^{n+1} =& \left(2- \frac 3 n \right) \psi_{\pm 1}^n + \left (\Delta t \right)^2  \left[ - H_0 \psi_{\pm 1}^n  -  q \psi_{\pm 1}^n   \right. \nonumber \\ 
&\left.  - \,  \delta \left( \lvert \psi_{\pm 1}^n \lvert ^2   + \lvert \psi_{0}^n \lvert ^2 - \lvert \psi_{\mp 1}^n \lvert ^2\right) \psi_{\pm 1}^n - \delta  \, \left(\psi_0^n \right)^2 \left({\psi_{\mp 1}^n}\right)^* \right. \nonumber \\
&\left.+ \, \mu \psi_{\pm 1}^n  \right]  
- \, \left(1 - \frac 3 n \right) \psi_{\pm 1}^{n-1},
\end{align}

 \begin{align}
\psi_{0}^{n+1} =& \left(2- \frac 3 n \right) \psi_{0}^n + \left (\Delta t \right)^2  \left[  -H_0 \psi_{0}^n- \delta \left( \lvert \psi_{1}^n \lvert ^2  + \lvert \psi_{-1}^n \lvert ^2\right) \psi_{0}^n \right. \nonumber \\ 
&\left.- \, 2  \delta \, \psi_{-1}^n \left(\psi_{ 0}^n\right)^* \psi_1^n + \mu \psi_{0}^n \right] 
- \left(1 - \frac 3 n \right) \psi_{0}^{n-1},
\end{align}
 where $ t = n \Delta t$ with time step $\Delta t$ and $n$ being the number of iterations made. Naturally, by the superscript we mean that
 $\psi_i^n=\psi_i(n \Delta t)=\psi_i(t)$.

Making use of gradient restarting, the ACTN scheme for the spin-1 system is given by
 
 \begin{align}
\psi_{\pm 1}^{n+1} =& \left(2- \frac 3 {\tilde{n}} \right) \psi_{\pm 1}^n + \left (\Delta t \right)^2  \left[ - H_0 \psi_{\pm 1}^n  -  q \psi_{\pm 1}^n   \right. \nonumber \\ 
&\left.  -\,  \delta \left( \lvert \psi_{\pm 1}^n \lvert ^2   + \lvert \psi_{0}^n \lvert ^2 - \lvert \psi_{\mp 1}^n \lvert ^2\right) \psi_{\pm 1}^n - \delta \, \left(\psi_0^n\right)^2 \left({\psi_{\mp 1}^n}\right)^* \right. \nonumber \\ 
&\left.+ \, \mu \psi_{\pm 1}^n  \right]   
- \,\left(1 - \frac 3 {\tilde{n}} \right) \psi_{\pm 1}^{n-1},
\end{align}

\begin{figure*}
\includegraphics{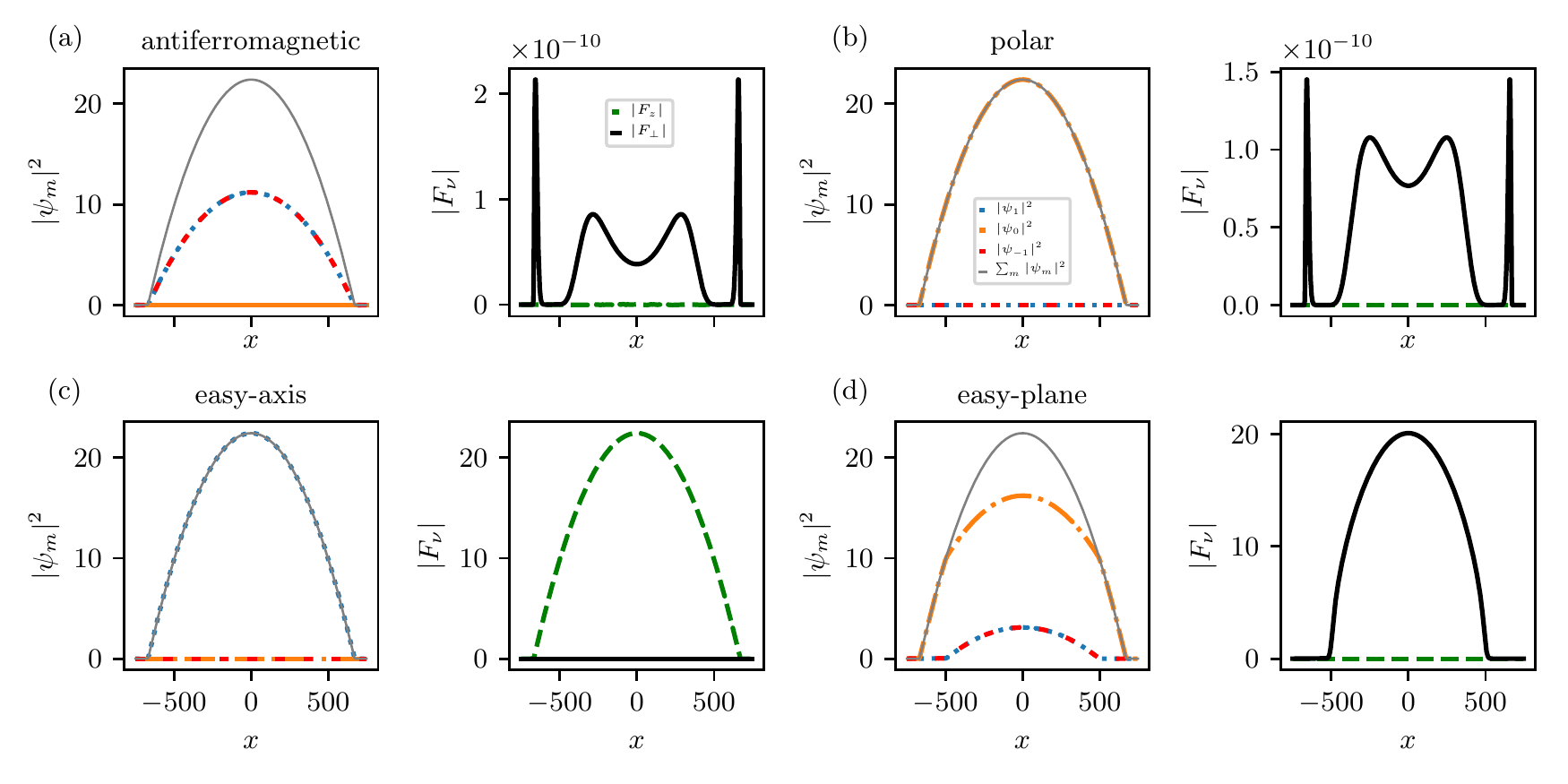}
\caption{\label{fig:NumGroundStates} Absolute value squared of the ground state wave functions, $\lvert \psi_{m} (x) \rvert^2$,  with $m =  0, \pm 1$ and the corresponding spin configurations, $\lvert F_{\nu} (x) \rvert$, with $\nu = z, \perp$ as a function of the spatial position $x$ obtained by means of the ACTN method for a trapped spin-1 Bose gas. 
The four panels correspond to values of $(\delta, q)$ of (a) $(5 \cdot 10^{-3}, -0.1)$, (b) $(5 \cdot 10^{-3}, 0.1)$, (c) $(-5 \cdot 10^{-3}, -0.1)$ and (d) $(-5 \cdot 10^{-3}, 0.1)$. 
The wave function of the $m_{\mathrm{F}} = 0$ component is depicted by a dash-dotted orange line. The $m_{\mathrm{F}} = \pm 1$ components are shown with blue dots and red dashes respectively.  The total density $\sum_m \lvert \psi_m (x) \rvert^2$ is illustrated by the grey solid line.
The amplitude of the transversal spin $\lvert F_{\perp} \rvert $ is given by the solid black line and the amplitude of the $F_z$-magnetization by the green dashes. 
The analysis is performed for parameters $\Omega = 10^{-2}$ and $N= 20000$ to mimic experimental settings. 
The associated phases of the spin-1 system are shown in the titles of the subplots. 
Note the scale of the amplitude of the spin in panels (a) and (b), which illustrates the numerical error arising from the error tolerance used for the ACTN scheme. }
\end{figure*}

 \begin{align}
\psi_{0}^{n+1} =& \left(2- \frac 3 {\tilde{n}} \right) \psi_{0}^n + \left (\Delta t \right)^2  \left[  -H_0 \psi_{0}^n- \delta \left( \lvert \psi_{1}^n \lvert ^2  + \lvert \psi_{-1}^n \lvert ^2\right) \psi_{0}^n \right. \nonumber \\ 
&\left.- \,  2  \delta \, \psi_{-1}^n \left(\psi_{ 0}^n\right)^* \psi_1^n + \mu \psi_{0}^n \right] 
- \left(1 - \frac 3 {\tilde{n}} \right) \psi_{0}^{n-1}.
\end{align}
Here, $\tilde{n}$ starts at 1 and is increased by 1 in each iteration step. $\tilde{n}$ is then reset to 1 when the (discretized in time) multicomponent
generalization of the gradient restarting condition for functionals
\begin{align}
0 < \, &\left <   - H_0 \psi_{1}^n  -  q \psi_{1}^n   -  \delta \left( \lvert \psi_{1}^n \lvert ^2   + \lvert \psi_{0}^n \lvert ^2 - \lvert \psi_{-1}^n \lvert ^2\right) \psi_{1}^n  \right.\nonumber \\ &\left. \,\,
- \, \delta \, \left(\psi_0^n\right)^2 \left({\psi_{-1}^n}\right)^*+ \mu \psi_{1}^n  \, , \, \psi_1^{n+1} - \psi_1^n \right >   \nonumber \\
&+ \,  \left <  -H_0 \psi_{0}^n- \delta \left( \lvert \psi_{1}^n \lvert ^2  + \lvert \psi_{-1}^n \lvert ^2\right) \psi_{0}^n \right. \nonumber \\ 
&\quad \left. \,\, - \,  2  \delta \, \psi_{-1}^n \left(\psi_{ 0}^n\right)^* \psi_1^n + \mu \psi_{0}^n  \, , \, \psi_0^{n+1} - \psi_0^n \right >    \nonumber \\
&+ \, \left <   - H_0 \psi_{- 1}^n  -  q \psi_{-1}^n   -  \delta \left( \lvert \psi_{-1}^n \lvert ^2   + \lvert \psi_{0}^n \lvert ^2 - \lvert \psi_{1}^n \lvert ^2\right) \psi_{-1}^n  \right.\nonumber \\ &\left. \,\,
\quad - \, \delta \,  \left(\psi_0^n\right)^2 \left({\psi_{1}^n}\right)^*+ \mu \psi_{- 1}^n  \, , \, \psi_{-1}^{n+1} - \psi_{-1}^n \right >   
\end{align}
is satisfied and $n > n_{\mathrm{res}}$ holds. 
Here, $\langle a \,, b \rangle= \sum_i a_i^* b_i$ denotes the
complex inner product.

To successfully apply the ACTN method a preconditioner has to be included. We choose the preconditioner to be $P = c - d^2/dx^2$, with the variable $x$ being the argument of the fields $\psi_m(x)$ 
and the constant $c$ being a real number. 
Due to the damping term in the ACTN scheme we additionally need to normalize 
the wave functions to the total particle number $N$ after each iteration step. 
The chemical potential $\mu$ is treated as a Lagrange multiplier and can be calculated from either Eqs.~(\ref{eq:EOM1TI}) or (\ref{eq:EOM2TI}). The full ACTN scheme with $\mu$  being calculated from Eq.~(\ref{eq:EOM2TI}) can be written as 

\begin{align}
\mu_n = \frac {\left <  -H_0 \psi_{0}^n- \delta \left( \lvert \psi_{1}^n \lvert ^2  + \lvert \psi_{-1}^n \lvert ^2\right) \psi_{0}^n - 2 \,   \delta \psi_{-1}^n \left(\psi_{ 0}^n\right)^* \psi_1^n \, , \,  P^{-1} \psi_0^n \right>} {\left < \psi_0^n \, , \, P^{-1} \psi_0^n \right>},
\end{align}

\begin{align}
\tilde{\psi}_{\pm 1}^{n+1} =& \left(2- \frac 3 {\tilde{n}} \right) \psi_{\pm 1}^n + \left (\Delta t \right)^2 P^{-1} \left[ - H_0 \psi_{\pm 1}^n  -  q \psi_{\pm 1}^n   \right. \nonumber \\ 
&\left.  - \,   \delta \left( \lvert \psi_{\pm 1}^n \lvert ^2   + \lvert \psi_{0}^n \lvert ^2 - \lvert \psi_{\mp 1}^n \lvert ^2\right) \psi_{\pm 1}^n - \delta \, \left(\psi_0^n\right)^2 \left({\psi_{\mp 1}^n}\right)^*  \right. \nonumber \\ 
&\left. + \mu_n \psi_{\pm 1}^n  \right]  - \, \left(1 - \frac 3 {\tilde{n}} \right) \psi_{\pm 1}^{n-1},
\end{align}

 \begin{align}
\tilde{\psi}_{0}^{n+1} =& \left(2- \frac 3 {\tilde{n}} \right) \psi_{0}^n + \left (\Delta t \right)^2  P^{-1} \left[  -H_0 \psi_{0}^n- \delta \left( \lvert \psi_{1}^n \lvert ^2  + \lvert \psi_{-1}^n \lvert ^2\right) \psi_{0}^n \right. \nonumber \\ 
&\left.- \, 2  \delta \,  \psi_{-1}^n \left(\psi_{ 0}^n\right)^* \psi_1^n + \mu_n \psi_{0}^n \right] 
-\left(1 - \frac 3 {\tilde{n}} \right) \psi_{0}^{n-1},
\end{align}

\begin{align}
\psi_{0, \pm 1}^{n+1} =  \frac {\tilde{\psi}_{0, \pm 1}^{n+1} \sqrt{N}}{\left [\left < \tilde{\psi}_1^{n+1} \,,\, \tilde{\psi}_1^{n+1} \right> + \left < \tilde{\psi}_0^{n+1} \,, \,\tilde{\psi}_0^{n+1} \right> + \left < \tilde{\psi}_{-1}^{n+1}\, ,\, \tilde{\psi}_{-1}^{n+1} \right>\right]^{1/2}} .
\end{align}
The convergence of the method depends on the choice of the constant $c$ in the preconditioner $P$, the time step $\Delta t$ as well as the minimum number of iterations that have to be performed before applying the gradient restarting $n_{\mathrm{res}}$. Note that the ACTN scheme can also be carried out in Fourier space which allows for a straightforward computation of the action
of the inverse of the preconditioner on Fourier modes $P^{-1} e^{i k x}= [1/(c + k^2)] e^{i k x}$.

\subsection{Numerical results}
\label{sec:NumResGS}

In this subsection we apply the ACTN method to map out the phase diagram for our trapped spin-1 Bose gas. 
Motivated by considerations of experimentally accessible
regimes \cite{PhysRevA.75.023617,PhysRevA.77.033612}, we choose $\lvert \delta \rvert = 5 \cdot 10^{-3}$, $\Omega = 10^{-2}$ and $N= 20000$. 
The numerics is performed on a one-dimensional grid with 512 grid points and the error tolerance of the ACTN is set to $10^{-10}$. 
The choice of parameters can correspond to a 1D condensate with peak density $n_{\mathrm{p}} \approx 93 \cdot 10^6 \mathrm{m}^{-1}$ confined in a trap with $\omega_{\perp}= 100 \, \omega_{\parallel} = 2 \pi \cdot 200$ Hz.

To provide a specific example, we now discuss the numerically obtained ground states for a quadratic Zeeman energy of $\lvert q \rvert = 0.1 $ (using the dimensionless units introduced in \Sect{Model'}).
Our initial guess of the wave function is a Gaussian, centered around the middle of the trap, with width $\sigma = 500/ \sqrt{2}$ in each of the $m_{\mathrm{F}}$ components. 
To converge to the ground state in the easy-axis phase we need to explicitly break the symmetry between the $m_{\mathrm{F}} = \pm 1$ components. 
As the equations of motion  are symmetric in the $m_{\mathrm{F}} = \pm 1$ components we have to impose a slight imbalance between them  in the initial wave function. 
We find that an imbalance of $0.2 \%$ is sufficient to let the system converge to either one or the other degenerate easy-axis ground state. 
Note that this does not affect the generality of the method to find ground states of the system without the need of an accurate initial guess.

The absolute value squared of the ground state wave functions and the corresponding spin configurations are depicted in Fig.~\ref{fig:NumGroundStates}.
We start by discussing the results for antiferromagnetic spin interactions which in our study corresponds to $\delta = 5 \cdot 10^{-3}$.

(i) \emph{polar phase} -- 
Convergence to the ground state at $q = 0.1$ within our preset tolerance is reached after $\simeq 700$
 iterations. 
The corresponding absolute value squared of the wave functions only being non-zero for the $m_{\mathrm{F}} = 0$ component as well as the vanishing spin (see Fig.~\ref{fig:NumGroundStates}(b)) clearly shows that the system is in the polar phase.
The chemical potential of the ground state is $\mu = 22.4$.
This value corresponds to $\mu = n_{\mathrm{p}} c_0$ obtained within the Thomas-Fermi approximation.

(ii) \emph{antiferromagnetic phase} -- 
In case of $q=-0.1$ the ACTN method needs $\simeq 600$ iterations to converge to the ground state. 
The data in Fig.~\ref{fig:NumGroundStates}(a), showing an equal non-zero absolute value squared of the wave functions of the $m_{\mathrm{F}} = \pm 1$ components and a vanishing spin, confirms that the system is in the antiferromagnetic phase.
The chemical potential here is $\mu =22.3$. 

For both settings (i) and (ii) we find that taking the parameters $\Delta t = 0.5$, $c=7$ and $n_{\mathrm{res}} = 50$ leads to an efficient convergence of the numerical scheme.

In the following we present the results obtained for ferromagnetic spin interactions which in our case is represented by $\delta = - 5 \cdot 10^{-3}$.

(iii) \emph{easy-axis phase} --  
At $q=-0.1$ we find two degenerate ground states after $\simeq 10000$ iterations.  The system is in the easy-axis phase which is validated by the non-zero $F_z$-magnetization (see Fig.~\ref{fig:NumGroundStates}(c)).
The chemical potential is $\mu =  22.23$. 
Efficient convergence of the ACTN is reached for parameters $\Delta t = 0.5$, $c=15$ and $n_{\mathrm{res}} = 100$. 

(iv) \emph{easy-plane phase} --  
At $q=0.1$ it takes $\simeq 2000$ iterations to converge to the ground state. The transversal spin depicted in  Fig.~\ref{fig:NumGroundStates}(d) clearly shows that the system is in the easy-plane phase.
The chemical potential is found to be $\mu = 22.38$. 
Taking the ACTN parameters to be $\Delta t = 0.5$, $c = 7$ and $n_{\mathrm{res}} = 200$ leads to an efficient convergence of the scheme in this case.

The zero-temperature phase transition
between the easy-plane and the polar phase occurs at $q = q_0$. 
In a homogeneous system described on the level of mean-field equations the transition is determined by $q_0 = 2 n_{\mathrm{h}} \lvert \delta \rvert$, where $n_{\mathrm{h}}$ is the homogeneous total density of the system. 
In a trapped system it is a priori not clear which density, if any,
might enter this type of critical-point relation. 

Using the ACTN method we are able to numerically determine the position of the phase transition within our mean-field approximation.
To do so we continuously increase the quadratic Zeeman energy starting at $q = 0$ and let the ACTN converge to the corresponding ground state. 
We then calculate the amplitude of the transverse spin $\lvert F_{\perp} (x) \rvert$ for the ground-state configuration. 
Crossing the phase transition the transverse spin should drop to zero as the system enters the unmagnetized polar phase. 
We define the phase transition to occur when $\lvert F_{\perp}  (x) \rvert_{\infty}  < 10^{-2}$, where $\lvert \, \cdot \, \rvert_{\infty}$ denotes the $L^{\infty}$ norm. 
We find that $q_{0} = 0.2236$ marks the phase transition in the trapped system.
This value is in good agreement with $q_0^{\mathrm{p}} = 2  n_{\mathrm{p}} \lvert \delta \rvert =0.224$ corresponding to the peak density.  
The position of the phase transition is thus determined by the peak density of the trapped system.
Note that the ACTN method needs $\simeq 4 \cdot 10^{5}$ iterations to converge to the ground state in the vicinity of the phase transition.
We hence observe that the number of iterations needed to converge to the ground state increases significantly close to the phase transition. Nevertheless, the method is still able to converge to the relevant ground state.

Choosing a value of $q > q_0$, i.e., being again in the polar phase, 
the same parameters $\Delta t$, $c$ and $n_{res}$ as for the polar phase with antiferromagnetic spin interactions can be used to achieve an efficient convergence of the ACTN scheme, i.e., the convergence of the numerical scheme is independent of the sign of $\delta$ as this term vanishes for an unmagnetized state.

\section{Stability analysis} \label{sec:Stability}

\begin{table*}
\scriptsize
\caption{\label{tab:OverviewBog} Stability properties of the spin-1 ground states derived within homogeneous Bogoliubov theory.
The abbreviations P, AF, EP, EA  stand for polar, antiferromagnetic, easy-plane, and easy-axis. 
Stable regimes of the listed excitation spectra are marked with S, unstable regimes with U. 
In case of an instability, the maximal growth rate $\gamma(\tilde{k})$, with $\tilde{k}$ being the respective  most unstable momentum mode, is stated. }
\begin{ruledtabular}
\begin{tabular}{cccccc}

 State& Energy spectrum& \multicolumn{4}{c}{Stability properties in \dots phase}\\
 && AF & P & EA & EP \\ \hline
 &&&&& \\
 P & $E_{\mathrm{ph}}^{\mathrm{P}} (k)= \sqrt{ \epsilon_k ( \epsilon_k +2 n_{\mathrm{h}} c_0)}$ &  S &S &S &S \\
 &&&&& \\
& $E_{\mathrm{s}} (k)= \sqrt{ \left (\epsilon_k + q \right) \left(\epsilon_k + q + 2 n_{\mathrm{h}} \delta  \right)}$ & U &S & U &U\\
&& $\gamma_{\mathrm{s}} (\tilde{k} ) = \left | \Im \left( \sqrt{ q \left(q + 2  n_{\mathrm{h}} \delta \right)} \right) \right |$ && $\gamma_{\mathrm{s}} (\tilde{k} ) =  n_{\mathrm{h}}| \delta |$&$\gamma_{\mathrm{s}} (\tilde{k} ) = \left | \Im \left( \sqrt{ q \left(q + 2  n_{\mathrm{h}} \delta \right)} \right) \right |$\\
&&  for $0 < -q <  n_{\mathrm{h}} \delta$ & && for $-n_{\mathrm{h}} \delta < q < -2 n_{\mathrm{h}} \delta$\\
&& $\gamma_{\mathrm{s}} (\tilde{k} ) =  n_{\mathrm{h}}| \delta |$ \,  for $n_{\mathrm{h}} \delta\leq -q$ &&& $\gamma_{\mathrm{s}} (\tilde{k} ) =  n_{\mathrm{h}}| \delta |$ \,  for $0 \leq q \leq -n_{\mathrm{h}}  \delta$\\
 &&&&& \\
AF & $E_{\mathrm{ph}}^{\mathrm{AF}} (k)= \sqrt{ \epsilon_k ( \epsilon_k +2 n_{\mathrm{h}} c_0)}$ & S &S &S &S\\
 &&&&& \\
& $E_{\mathrm{m}} (k)= \sqrt{ \epsilon_k ( \epsilon_k +2 n_{\mathrm{h}} \delta)}$ & S & S ($\delta >0$), U  ($\delta <0$) & U & U \\ 
&&& $\gamma_{\mathrm{m}} (\tilde{k})= n_{\mathrm{h}} | \delta |$ & $\gamma_{\mathrm{m}} (\tilde{k})= n_{\mathrm{h}} | \delta |$ & $\gamma_{\mathrm{m}} (\tilde{k})= n_{\mathrm{h}} | \delta |$\\ 
 &&&&& \\
& $E_{\mathrm{g}} (k)= \sqrt{ \left(\epsilon_k -q \right)^2 + 2 n_{\mathrm{h}} \delta \left(\epsilon_k - q\right)}$ & S & U &S ($q <  2 n_{\mathrm{h}} \delta$) , U ($q >  2 n_{\mathrm{h}} \delta$) & U \\
&&& $\gamma_{\mathrm{g}} (\tilde{k} )= \left | \Im \left (\sqrt{q \left ( q- 2 n_{\mathrm{h}} \delta \right)} \right) \right |$ &  $\gamma_{\mathrm{g}} (\tilde{k} )=  n_{\mathrm{h}} |\delta |$ & $\gamma_{\mathrm{g}} (\tilde{k} )=  n_{\mathrm{h}} |\delta |$ \\ 
&&& for $q < n_{\mathrm{h}} \delta$ and $\delta > 0$ &for $n_{\mathrm{h}} \delta < q < 0$&\\
&&& $\gamma_{\mathrm{g}} (\tilde{k} )=  n_{\mathrm{h}} |\delta |$ &$\gamma_{\mathrm{g}} (\tilde{k} )= \left | \Im \left (\sqrt{q \left ( q- 2 n_{\mathrm{h}} \delta \right)} \right) \right |$& \\
&&&  for $q \geq n_{\mathrm{h}} \delta$ and $\delta > 0$ &for $2 n_{\mathrm{h}} \delta < q  \leq n_{\mathrm{h}} \delta$&\\
&&&&& \\
EP & $E_{0} = \sqrt{\epsilon_k \left(\epsilon_k + q \right)}$ & U & S & U & S \\
&& $\gamma_0 (\tilde{k} )=| q| /2$ && $\gamma_0 (\tilde{k} )= | q| /2$  &\\
&&&&& \\
EA &  $E_{\mathrm{ph}}^{\mathrm{EA}} = \sqrt{\epsilon_k \left[ \epsilon_k + 2\left (1 + \delta \right)  n_{\mathrm{h}} \right]}$ &S &S& S ($\delta \geq -1$) , U ($\delta < -1$) & S ($\delta \geq -1$) , U ($\delta < -1$)  \\ 
&&&&
$\gamma_{\mathrm{ph}}^{\mathrm{EA}}(\tilde{k} ) = \sqrt{3} \left | (1+\delta)  \right | n_{\mathrm{h}}$ & 
$\gamma_{\mathrm{ph}}^{\mathrm{EA}}(\tilde{k} ) = \sqrt{3} \left | (1+\delta)  \right | n_{\mathrm{h}}$
\end{tabular}
\end{ruledtabular}
\end{table*}

In this section we perform the stability analysis for the spin-1 ground states throughout the $(\delta,q)$-plane of the spin-1 
phase diagram. 
The stability properties are extracted by numerically solving the Bogoliubov de-Gennes (BdG) equations in presence of a trapping
potential. 
The BdG equations are obtained by considering small perturbations about a possible stationary state of the system to linear order.

We first review Bogoliubov theory in a homogeneous spin-1 system for later comparison with the numerical results for the trapped setup.
Solving the BdG equations describing the linear excitations about a particular stationary state requires to first determine the wave functions of this state.
To find the desired stationary state for any parameter set $(\delta,q)$ we employ a highly accurate Newton scheme, which we introduce
in the second part of this section.
We then derive the BdG equations for the one-dimensional trapped spin-1 Bose gas.
Finally we discuss the numerically obtained stability properties of the spin-1 ground states.

\subsection{Bogoliubov excitations in a homogeneous system}
\label{sec:BogHomSystem}

In the following we give a brief summary of the Bogoliubov theory for the spin-1 ground states in absence of a trapping potential ($\Omega = 0$). 
The results are summarized in \Tab{OverviewBog}.
The Bogoliubov excitation spectra enable us to determine the dynamical stability of the ground states in different phases. Whenever the mode energies become imaginary a dynamical instability occurs as relevant momentum modes will grow exponentially in time. 
In that case the growth rates of the unstable modes can be calculated from the excitation spectra (i.e., from the imaginary parts of the corresponding
eigenfrequencies). 
We will use the Bogoliubov predictions made for the homogeneous system for later comparison to numerical results obtained for the trapped system with $\Omega \neq 0$.   
A detailed analysis of the homogeneous theory can be found in \cite{Kawaguchi2012a.PhyRep.520.253}.

\subsubsection{Excitations about the polar state}
\label{sec:BogHomPolarState}

Diagonalizing the Bogoliubov Hamiltonian for a small perturbation about the polar state $\psi \sim (0,1,0)^T$ in a homogeneous spin-1 system one obtains one phonon mode and two modes corresponding to excitations in the transverse spin direction. 
The spectrum of the phonon mode is given by 
\begin{equation}
E_{\mathrm{ph}}^{\mathrm{P}} (k)= \sqrt{ \epsilon_k ( \epsilon_k +2 n_{\mathrm{h}} c_0)},
\end{equation} 
with $\epsilon_k = k^2/2$.
This mode is stable irrespective of the parameters $q$ and $\delta$. 
The spectrum of the transverse spin excitations reads

\begin{equation}
\label{eq:DispersionPolar}
E_{\mathrm{s}} (k)= \sqrt{ \left (\epsilon_k + q \right) \left(\epsilon_k + q + 2 n_{\mathrm{h}} \delta  \right)}.
\end{equation}

This mode is dynamically unstable whenever the parameters $q$ and $\delta$ are chosen in a way that the expression under the square root becomes negative.

(i) \emph{antiferromagnetic phase} --
In case of $q<0$ and $\delta>0$, i.e., in the antiferromagnetic phase, three different instability regimes exist. 
For $0 < -q <  n_{\mathrm{h}} \delta$ momentum modes up to an ultra-violet (UV) cutoff $k_{\mathrm{UV}} = \sqrt{-2 q}$ are unstable. 
The most unstable mode is $\tilde{k} =0$ with growth rate 
\begin{equation}
\label{eq:gammafPolar1}
\gamma_{\mathrm{s}} \left(\tilde{k} \right) = \left | \Im \left(E_{\mathrm{s}} (0) \right)  \right | = \left | \Im \left( \sqrt{ q \left(q + 2  n_{\mathrm{h}} \delta \right)} \right) \right |,
\end{equation}
where the symbol $\Im$ denotes the imaginary part of a complex number. 
For $n_{\mathrm{h}} \delta\leq -q$ the most unstable mode is $\tilde{k} = \sqrt{-2 (q + n_{\mathrm{h}} \delta)}$ with growth rate 
\begin{equation}
\label{eq:gammafPolar2}
\gamma_{\mathrm{s}} \left(\tilde{k} \right) =  n_{\mathrm{h}} \left | \delta \right |.
\end{equation}
In case of  $2 n_{\mathrm{h}} \delta < -q$ an additional infra-red (IR) cutoff of the instability region occurs at $k_{\mathrm{IR}} = \sqrt{- (q + 2  n_{\mathrm{h}} \delta)}$. 

The same scenario is present for $\delta < 0$. 

(ii) \emph{easy-plane phase} --
For $0 < q < -2 n_{\mathrm{h}} \delta$, i.e, in the easy-plane phase, the first two of the above stated instability regimes can be found. 
In case of $-n_{\mathrm{h}} \delta < q < -2 n_{\mathrm{h}} \delta$ the most unstable mode is $\tilde{k}=0$ with growth rate given by Eq.~(\ref{eq:gammafPolar1}). 
For parameters $0 \leq q \leq -n_{\mathrm{h}}  \delta$ the most unstable mode occurs at $\tilde{k} = \sqrt{-2 (q + n_{\mathrm{h}} \delta)}$. 
The corresponding growth rate is stated in Eq.~(\ref{eq:gammafPolar2}). 

(iii) \emph{easy-axis phase} --
Moving to $q<0$, i.e., entering the easy-axis phase, the additional IR cutoff of the instability region occurs as mentioned above. The most unstable mode is $\tilde{k} = \sqrt{-2 (q + n_{\mathrm{h}} \delta)}$ with growth rate given by Eq.~(\ref{eq:gammafPolar2}).

\subsubsection{Excitations about the antiferromagnetic state}
\label{sec:BogHomAFState}

Diagonalizing the Bogoliubov Hamiltonian for a small perturbation about the antiferromagnetic state $\psi \sim (1,0,1)^T$ in a homogeneous spin-1 system one obtains one stable uncoupled phonon mode, given by 
\begin{equation}
E_{\mathrm{ph}}^{\mathrm{AF}} (k)= \sqrt{ \epsilon_k ( \epsilon_k +2 n_{\mathrm{h}} c_0)}. 
\end{equation}
In addition, an uncoupled magnon mode with spectrum
\begin{equation}
E_{\mathrm{m}} (k)= \sqrt{ \epsilon_k ( \epsilon_k +2 n_{\mathrm{h}} \delta)}
\end{equation}
exists. The magnon mode exhibits unstable momentum modes up to a UV cutoff of $k_{\mathrm{UV}} = \sqrt{-4 n_{\mathrm{h}} \delta}$ in case of $\delta <0$.
Irrespective of the parameter $q$, the most unstable mode is $\tilde{k} = \sqrt{-2 n_{\mathrm{h}} \delta}$ with corresponding growth rate
\begin{equation}
\label{eq:gammamagAF}
\gamma_{\mathrm{m}} \left(\tilde{k} \right)= n_{\mathrm{h}} \left | \delta \right |.
\end{equation}
Hence one finds unstable modes showing the same maximal growth rate within the polar, easy-plane and easy-axis phase associated with 
ferromagnetic spin interactions.

Furthermore, a quadratic mode described by
\begin{equation}
E_{\mathrm{g}} (k)= \sqrt{ \left(\epsilon_k -q \right)^2 + 2 n_{\mathrm{h}} \delta \left(\epsilon_k - q\right)}
\end{equation}
is present in the system. 

(i) \emph{polar phase} --
For $\delta>0$ and $q>0$, i.e., in the polar phase for antiferromagnetic spin interactions, we find two different instability regimes. 
In case of $q < n_{\mathrm{h}} \delta$ the most unstable mode occurs at $\tilde{k}=0$. Its growth rate reads 
\begin{equation}
\label{eq:gamma0AF1}
\gamma_{\mathrm{g}} \left(\tilde{k} \right)= \left | \Im \left (\sqrt{q \left ( q- 2 n_{\mathrm{h}} \delta \right)} \right) \right |.
\end{equation}
The second regime emerges for $q \geq n_{\mathrm{h}} \delta$ where the most unstable mode becomes $\tilde{k} = \sqrt{2 (q- n_{\mathrm{h}} \delta)}$. The growth rate is given by
\begin{equation}
\label{eq:gamma0AF2}
\gamma_{\mathrm{g}} \left(\tilde{k} \right)=  n_{\mathrm{h}} \left |\delta \right |.
\end{equation}
For $\delta < 0$ and $q > - 2n_{\mathrm{h}} \delta$, i.e., in the polar phase for ferromagnetic spin interactions, the quadratic mode
is stable.

(ii) \emph{easy-plane and easy-axis phase} --
For $\delta < 0$ and $n_{\mathrm{h}} \delta < q < -2 n_{\mathrm{h}} \delta$, i.e.~in the easy-plane phase and parts of the easy-axis phase, the most unstable mode is $\tilde{k} = \sqrt{2 (q- n_{\mathrm{h}} \delta)}$ with growth rate according to Eq.~(\ref{eq:gamma0AF2}). 
A second instability regime occurs within the easy-axis phase for $2 n_{\mathrm{h}} \delta < q  \leq n_{\mathrm{h}} \delta$. 
Here, $\tilde{k} =0$ is the most unstable mode with growth rate given by Eq.~(\ref{eq:gamma0AF1}). 
For values $q < 2 n_{\mathrm{h}} \delta$ the quadratic mode is dynamically stable. 

\subsubsection{Excitations about the easy-plane state}
\label{sec:BogHomEPState}

Diagonalizing the Bogoliubov Hamiltonian for a small perturbation about the easy-plane state $\psi \sim \left (\sqrt{1-q/q_0}, \sqrt{2(1 +q/q_0)}, \sqrt{1 -q/q_0} \right)^T$ in a homogeneous spin-1 system one obtains one gapless mode given by
\begin{equation}
E_{0} = \sqrt{\epsilon_k \left(\epsilon_k + q \right)}.
\end{equation}
This mode is dynamically unstable for $q<0$ irrespective of the spin interaction $\delta$. The most unstable mode is $\tilde{k} = \sqrt{-q}$ with growth rate
\begin{equation}
\label{eq:gamm0EP}
\gamma_0 \left(\tilde{k} \right)= \frac {\left | q \right |}2. 
\end{equation}
Note that there are two further modes that will not be discussed here.

\subsubsection{Excitations about the easy-axis state}
\label{sec:BogHomEAState}

Diagonalizing the Bogoliubov Hamiltonian for a small perturbation about the easy-axis state $\psi \sim (1,0,0)^T$  or  $\psi \sim (0,0,1)^T $ respectively in a homogeneous spin-1 system one obtains two single-particle like modes which are stable. 
The system exhibits an additional phonon mode with spectrum
\begin{equation}
E_{\mathrm{ph}}^{\mathrm{EA}} = \sqrt{\epsilon_k \left[ \epsilon_k + 2\left (1 + \delta \right)  n_{\mathrm{h}} \right]}.
\end{equation}
This mode is dynamically unstable for $\delta < -1$. 
The most unstable mode is $\tilde{k} = \sqrt{-2(1+\delta)n_{\mathrm{h}}}$ with growth rate given by 
\begin{equation}
\gamma_{\mathrm{ph}}^{\mathrm{EA}}\left(\tilde{k} \right) = \sqrt{3} \left | (1+\delta)  \right | n_{\mathrm{h}}.
\end{equation}
As we are studying experimentally realistic parametric regimes, the spin coupling $\lvert \delta \rvert $ is on the order of $\sim 10^{-2}$ so we expect to find no dynamically unstable modes for the easy-axis state irrespective of the parameter $q$ and the sign of $\delta$.

\subsection{Determining the wave functions of stationary states within the $(\delta,q)$-plane using an exact Newton method}
\label{sec:Newton}

Performing the stability analysis of the different states discussed
above within the $(\delta,q)$-plane of the spin-1 phase diagram, i.e., especially in regions where they are not the ground state anymore, requires to first numerically determine their wave functions for any given set of parameters $(\delta,q)$.
To achieve this goal, we employ an exact Newton method which is also capable of converging to excited states of the system. 
However, as highlighted above, this requires an initial guess for the wave function which is close to the desired state.
The Newton scheme for the spin-1 system can be cast into the form of a six-dimensional matrix equation:

\begin{equation}
\label{eq:NewtonScheme}
J \Delta \psi = \mathcal{F},
\end{equation}
where $\mathcal{F} = ( \mathcal{F}_1,\mathcal{F}_0,\mathcal{F}_{-1},\mathcal{F}_1^* ,\mathcal{F}_0^*,\mathcal{F}_{-1}^*)^T $ is a vector that contains the time-independent equations of motion (see Eqs.~(\ref{eq:EOM1TI}) and (\ref{eq:EOM2TI})) as well as their complex conjugated versions and  $\Delta \psi$ gives the correction to the wave function of the previous iteration of the Newton scheme with $\psi = ( \psi_1,\psi_0,\psi_{-1}, \psi_1^*, \psi_0^* , \psi_{-1}^* )^T$ being a vector of all spinor fields.
The Jacobian $J$ is given by the matrix
\begin{equation}
\label{eq:Jacobian}
J_{ij} = \frac {\partial \mathcal{F}_i} {\partial \psi_j},
\end{equation}
where $i,j \, \epsilon \,  \{0, \dots,5\}$ and the partial derivative is evaluated at the current wave function $\psi$. Note that we end up with a $6 N_{\mathrm{g}} \, \times  \, 6 N_{\mathrm{g}} $ matrix when taking $N_{\mathrm{g}}$ grid points to discretize the wave functions.

As we wish to converge to a state with fixed particle number we introduce a Lagrange multiplier $\lambda$ for the chemical potential $\mu$. 
This adds the following constraint to our Newton scheme
\begin{equation}
\mathcal{F}_{\lambda} \equiv  \int \left ( \lvert \psi_1 \rvert^2 +\lvert \psi_0 \rvert^2 + \lvert \psi_{-1} \rvert^2 \right) \mathrm{d}x - N = 0.
\end{equation}
Consequently, we get an additional row and column in the Jacobian such that we are dealing with $6 N_{\mathrm{g}} +1$ equations in the Newton scheme.
The modified scheme can be written as
\begin{equation}
\label{eq:NewtonSchemeWithLagrangeMult}
\tilde{J} \Delta \tilde{\psi} = \tilde{\mathcal{F}},
\end{equation}
with $\tilde{\psi} = ( \psi_1,\psi_0,\psi_{-1}, \psi_1^*, \psi_0^* , \psi_{-1}^*, \lambda )^T$. Note that all $\psi_m^{(*)}$ are vectors containing the wave function at grid points $1, \dots , N_{\mathrm{g}}$.
In each iteration step we calculate $\tilde{\mathcal{F}}$ and evaluate the Jacobian $\tilde{J}$ of the system.
The second derivative occurring in the equations of motion is obtained by means of a second-order center difference scheme.
By solving the eigenvalue equation (\ref{eq:NewtonSchemeWithLagrangeMult}) we obtain the correction to the wave function $\Delta \tilde{\psi}$. 
The Newton scheme terminates if the correction is smaller than a preset tolerance.

\subsection{Bogoliubov de-Gennes equations}
\label{sec:BdG}

The stability properties of a specific stationary state are deduced from numerically solving the corresponding Bogoliubov de-Gennes (BdG) equations.
In this subsection we present the derivation of the BdG equations for the trapped spin-1 system and elaborate on how to subsequently solve them.

As a first step we have to linearize the equations of motion about the stationary state
of interest. Thus we take the ansatz
\begin{equation}
\psi_m (x,t) = \left[ \Phi_{m}(x) + \epsilon \delta \psi_m (x,t) \right] e^{- i \mu t},
\end{equation}
with $m = 0, \pm 1$ labeling the three hyperfine components and $\Phi_{m}(x)$ being the wave function of each component at the stationary state; $\mu$ is the corresponding chemical potential; 
$\epsilon$ is a (formal) small parameter with $\epsilon \ll 1$ and $\delta \psi_m$ is the perturbation about the stationary state. 
Plugging this ansatz into Eqs.~(\ref{eq:EOM1Dimless}) and (\ref{eq:EOM2Dimless}) we obtain 
\begin{align}
\label{eq:EOMDeltaPsi}
i \epsilon \partial_t \delta \psi_m = \mathcal{F}_m &\left(\Phi_1 + \epsilon \delta \psi_1 , \Phi_0 + \epsilon \delta \psi_0 , \Phi_{-1} + \epsilon \delta \psi_{-1} , \right. \nonumber \\
& \, \, \left. \Phi_1^* + \epsilon \delta \psi_1^* , \Phi_0^* + \epsilon \delta \psi_0^* , \Phi_{-1}^* + \epsilon \delta \psi_{-1}^* \right ).
\end{align}
Here, the $\mathcal{F}_m$ are the functions introduced in Eqs.~(\ref{eq:EOM1TI}) and (\ref{eq:EOM2TI}).

Linearization of the equations of motion (\ref{eq:EOMDeltaPsi}) boils down to a Taylor expansion of $\mathcal{F}_m$ to first order in $\epsilon$. 
The expansion for $\mathcal{F}_m$ reads
\begin{align}
	\mathcal{F}_m \left(\dots \right) =& \,\, \mathcal{F}_m \left( \Phi  \right) + \epsilon \left \{ \left(\frac {\partial   \mathcal{F}_m }{\partial \Phi_1}\right)_{| \Phi} \delta \psi_1  +    \left( \frac {\partial   \mathcal{F}_m }{\partial \Phi_0} \right)_{| \Phi}  \delta \psi_0  \right. \nonumber \\
	&\left.  +  \left(\frac {\partial   \mathcal{F}_m }{\partial \Phi_{-1}} \right)_{| \Phi}  \delta \psi_{-1}   +  \left( \frac {\partial   \mathcal{F}_m}{\partial \Phi_1^*} \right)_{| \Phi}  \delta \psi_1^*  +  \left(\frac {\partial   \mathcal{F}_m }{\partial \Phi_0^*} \right)_{| \Phi}  \delta \psi_0^*  \right. \nonumber \\
	&\left. +  \left(\frac {\partial   \mathcal{F}_m }{\partial \Phi_{-1}^*} \right)_{| \Phi}  \delta \psi_{-1}^*     \right \} + \mathcal{O} \left(\epsilon^2\right).
\end{align}
Here, $\Phi = (\Phi_1 ,\Phi_0 ,\Phi_{-1} ,\Phi_1^* ,\Phi_0^* ,\Phi_{-1}^* )$ is a vector containing the wave functions at the stationary state. 
Note that $\mathcal{F}_m \left( \Phi \right) = 0$ for all components as $\Phi$ is a stationary state of the system. 
The partial derivatives of $\mathcal{F}_m$ are taken with respect to the stationary fields and are then evaluated at $\Phi$. 
To order $\epsilon$ we thus obtain
\begin{align}
\label{eq:EOMOrderEpsilon}
	 i  \partial_t \delta \psi_m =& \,\,\left(\frac {\partial   \mathcal{F}_m }{\partial \Phi_1}\right)_{| \Phi} \delta \psi_1  +    \left( \frac {\partial   \mathcal{F}_m }{\partial \Phi_0} \right)_{| \Phi}  \delta \psi_0  +  \left(\frac {\partial   \mathcal{F}_m }{\partial \Phi_{-1}} \right)_{| \Phi}  \delta \psi_{-1}    \nonumber \\
	&+  \left( \frac {\partial   \mathcal{F}_m}{\partial \Phi_1^*} \right)_{| \Phi}  \delta \psi_1^*  +  \left(\frac {\partial   \mathcal{F}_m }{\partial \Phi_0^*} \right)_{| \Phi}  \delta \psi_0^*  +  \left(\frac {\partial   \mathcal{F}_m }{\partial \Phi_{-1}^*} \right)_{| \Phi}  \delta \psi_{-1}^* .
\end{align}
To solve the BdG equations we make use of the ansatz
\begin{equation}
\label{eq:AnsatzModeFunc}
\delta \psi_m (x,t) = \left( u_m (x) e^{- i \omega t} + v_m^* (x) e^{i \omega^* t} \right),
\end{equation}
with mode functions $u_m , v_m$  and mode frequency $\omega$. 
Inserting the ansatz into Eq.~(\ref{eq:EOMOrderEpsilon}) and matching the phase factors to obtain a time-independent description, we end up with a system of six coupled equations. 
We can write the BdG equations as an eigenvalue problem of the form
\begin{equation}
\label{eq:BdGEquations}
\bar{J} \mathcal{M} = - \omega \mathcal{M}.
\end{equation}
Here, $\mathcal{M} = (u_1,u_0,u_{-1},v_1,v_0,v_{-1} )^T$ is a vector that contains all eigenmodes of the system. 
Note again that the vector has $6 N_{\mathrm{g}}$ entries after the discretization on a grid with $N_{\mathrm{g}}$ grid points.
The matrix $\bar{J}$ turns out to be the Jacobian introduced in Eq.~(\ref{eq:Jacobian}) whose lower
half of entries is multiplied by a factor of $-1$. We can formally write it as

\begin{equation}
\bar{J}_{ij} = \left[1 -2 \Theta(i-3) \right] \left(J_{ij} \right)_{| \Phi} ,
\end{equation}
where $i,j \, \epsilon \, \{0,\dots,5\}$ and the Heaviside theta function $\Theta$ is defined as $\Theta (z) = 1$ for $z \geq 0$.

The mode frequencies $\omega$ correspond to the eigenvalues of $\bar{J}$ and the mode functions $u_m, v_m$ are given by the eigenvectors.
We numerically solve the eigenvalue problem in Eq.~(\ref{eq:BdGEquations}) using the standard $_{-}$geev LAPACK routines in python.
Eigenmodes corresponding to mode frequencies with a non-zero imaginary part are dynamically unstable as they grow in time. Their growth rate is given by the magnitude of the imaginary part.

\subsection{Numerical results}
\label{sec:NumBdG}

In the following we discuss the stability properties of the spin-1 ground states in presence of a trapping potential.
In Sects.~\sect{PolarState}-\sect{EAState}, we investigate those properties for the polar, antiferromagnetic, easy-plane
and easy-axis state, respectively. 
The numerical settings are taken to be the same as in \Sect{NumResGS}, i.e., we take parameters $\lvert \delta \rvert = 5 \cdot 10^{-3}$ for the spin coupling and $\Omega = 10^{-2}$ for the normalized trap strength. 
We finally study the dependence of the stability properties on the strength of the spin coupling $\delta$ and the normalized  trap strength $\Omega$ for the example case of the polar state in Sects.~\sect{RoleOfDelta} and \sect{RoleOfOmega}. 

\subsubsection{Excitations about the polar state}
\label{sec:PolarState}

We investigate the dynamical stability of the polar state $\psi \sim \left(0,1,0 \right)^T$ throughout the different phases of our trapped spin-1 system. 
To get an overview of the stability properties of the polar state we
discuss results of the BdG analysis obtained for
the case example of $\lvert q \rvert = 0.1$. 
The initial guess for the wave functions used in the Newton scheme is taken to be a Gaussian, centered around the middle of the trap, with width $\sigma = 500 / \sqrt{2}$ in the $m_{\mathrm{F}} = 0$ component and $0$ in the $m_{\mathrm{F}} = \pm 1$ components. 
The Newton method converges to the polar state in any of the phases within 10 iterations when setting the error tolerance to $10^{-10}$. Due to a finite accuracy of the eigenvalue solver we only consider 
eigenmodes with imaginary part larger than $10^{-4}$ as modes with non-zero imaginary part.

\begin{figure}
\includegraphics{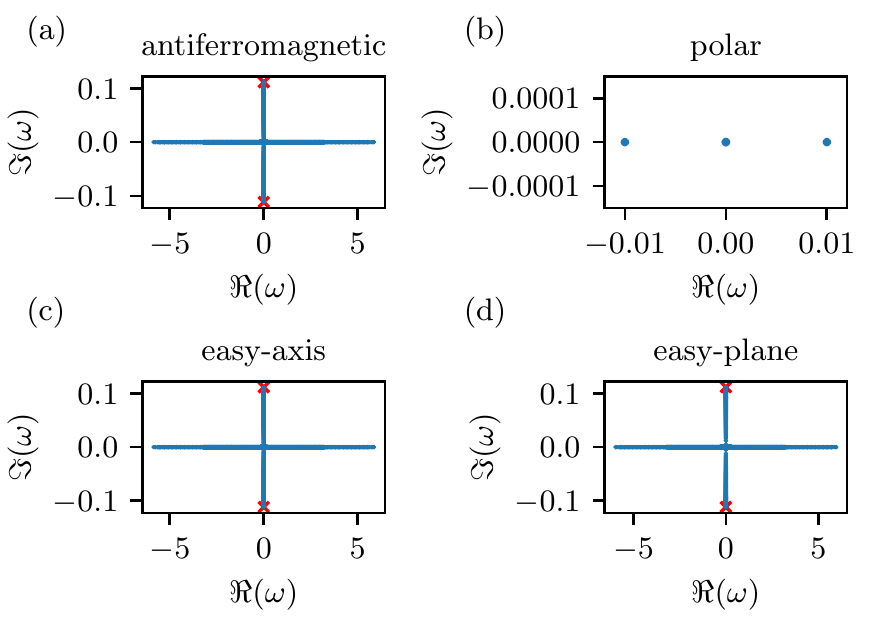}
\caption{\label{fig:BdGPolarState} Real ($\Re$) and imaginary ($\Im$) parts of the mode frequencies $\omega$
  resulting from the BdG analysis of the \emph{polar} state $\psi \sim \left(0,1,0 \right)^T$ within the (a) antiferromagnetic $(\delta, q) = ( 5\cdot 10^{-3} , -0.1)$, (b) polar $( 5\cdot 10^{-3} , 0.1)$, (c) easy-axis $( -5\cdot 10^{-3} , -0.1)$ and (d) easy-plane $(- 5\cdot 10^{-3} , 0.1)$ phase. 
In panels (a), (c) and (d) we see a continuum band of mode frequencies along the real and the imaginary axis indicating that the frequencies are either purely real or purely imaginary.
The red crosses mark the predicted mode frequency with the largest imaginary part within the given parameter regime derived from homogeneous Bogoliubov theory  by replacing the homogeneous density by the peak density of the trapped system. 
The prediction is in good agreement with the largest imaginary part of the numerically obtained mode frequencies. 
In panel (b) we only show the neutral and the dipolar mode of the system. 
Their appearance as the lowest eigenmodes confirms that our BdG method is working properly.}
\end{figure}

\begin{figure}
\includegraphics{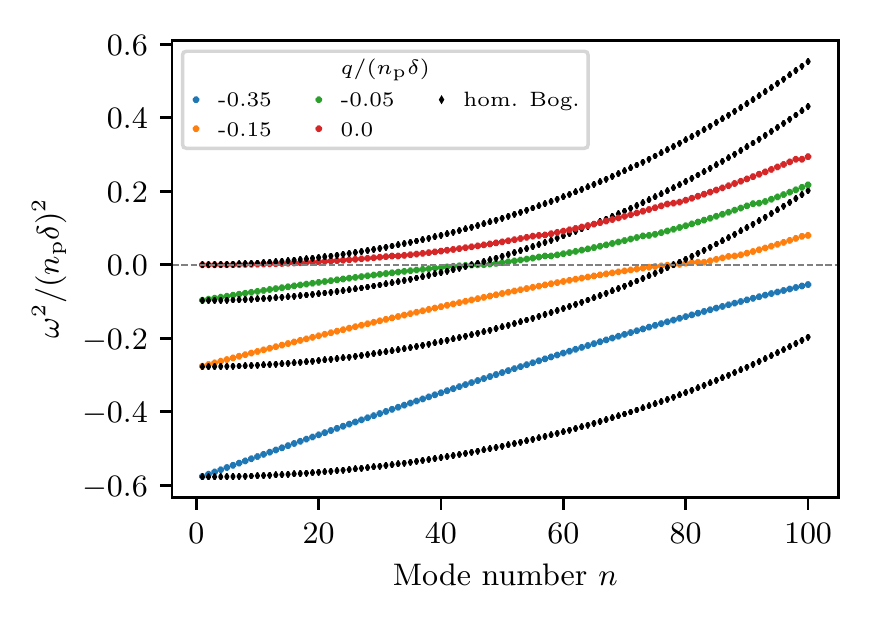}
\caption{\label{fig:EnergySpectrumPolarInAF}
Squared mode frequencies $\omega^2$ as a function of the mode number $n$ for the trapped system (colored dots) as compared to the homogeneous setting (black diamonds). 
The depicted data points for the trapped case are obtained by means of the BdG analysis of the \emph{polar} state using the parameters $-0.35 \leq q/ n_{\mathrm{p}} \delta \leq 0$, $\delta = 5 \cdot 10^{-3}$ and $\Omega = 10^{-2}$.
Data points for the homogeneous case result from \Eq{DispersionPolar}  using momenta $k_n = \pi n / L_\mathrm{b}$ associated with the $n$-th eigenmode in a one-dimensional box of length $L_\mathrm{b} = 2 \, R_{\mathrm{TF}}$, where $R_{\mathrm{TF}}$ is the Thomas-Fermi radius of the corresponding trapped system.
The presence of the trap leads to a reduction of the growth rates for all unstable modes except the most unstable one. 
However, it appears to have no effect on the crossing point to the stable regime. 
The mode frequencies are given in units of $n_\mathrm{p} \delta$. 
The grey dashed line marks the transition between the unstable ($\omega^2 < 0$) and the stable ($\omega^2 >0$) regime of modes.} 
\end{figure}

\begin{figure}
\includegraphics{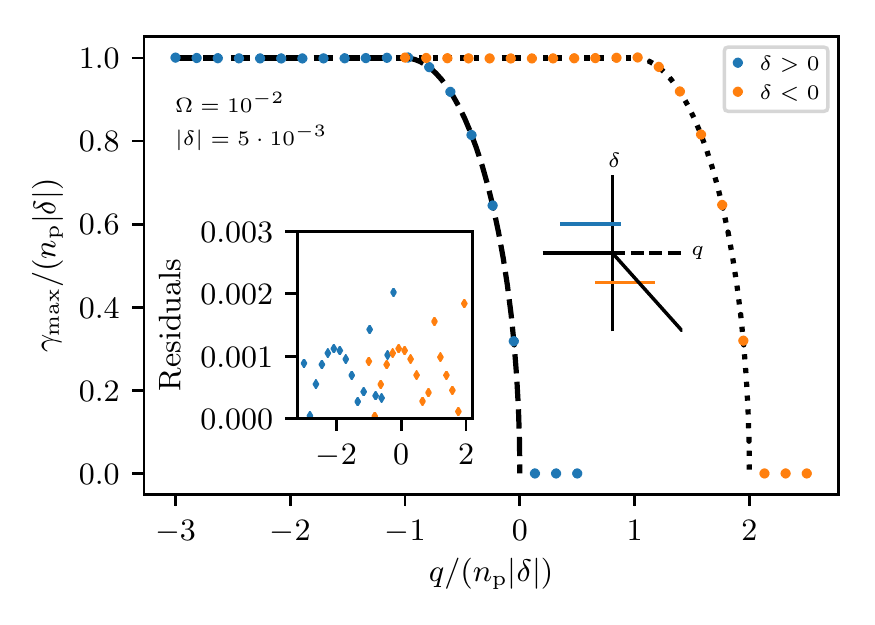}
\caption{\label{fig:MaxGrowthRatesPolarState} Maximal growth rate $\gamma_{\mathrm{max}} = \mathrm{max} \, | \Im (\omega ) |$ of the eigenmodes obtained by means of the BdG analysis of the \emph{polar} state  as a function of $q$ and the sign of $\delta$. 
The investigated parameter regime is indicated by the blue and orange solid lines in the schematic representation of the spin-1 phase diagram (c.f.~\Fig{PhaseDiag}).  
The analysis is performed for parameters $\Omega = 10^{-2}$ and $\lvert \delta \rvert = 5  \cdot 10^{-3}$. 
The growth rate as well as the quadratic Zeeman energy are given in units of $n_{\mathrm{p}} \lvert  \delta \rvert$.  For $\delta > 0$ the growth rate follows the homogeneous prediction (dashed line) for the whole parameter range. 
The same feature is found in case of $\delta < 0$ where the data agrees with the dotted lines.
The inset shows the residuals $|\gamma_{\mathrm{max, h}} - \gamma_{\mathrm{max}}|/ \gamma_{\mathrm{max,h}}$, with $\gamma_{\mathrm{max,h}}$ being the growth rate calculated from the homogeneous prediction. 
The deviation of the numerically extracted growth rates from the homogeneous prediction is less than $0.2 \%$ for all parameters considered. The color coding is as in the main frame.}
\end{figure}

(i) \emph{polar phase} --
To check whether our numerical BdG analysis is working properly we first study the stability of the polar state inside the polar phase, i.e., at the parameter
pair $(\delta, q) = (5 \cdot 10^{-3} , 0.1)$. 
As the polar state is the ground state in this phase it has to be stable. 
This corresponds to all mode frequencies $\omega$ being real. 
Performing the BdG analysis we find the imaginary parts of all obtained eigenmodes to be zero within our tolerance.
This confirms the expected stability of the polar state within the polar phase.
Fig.~\ref{fig:BdGPolarState}(b) shows the imaginary ($\Im (\omega)$) and real ($\Re (\omega)$) part of the two energetically lowest eigenmodes. 
The mode with eigenvalue $\omega = 0$ is the neutral mode.  
In addition we observe that the first mode on the real axis is located at $\lvert \omega \rvert = 0.01$. 
This mode is called the dipolar mode with mode frequency given by the normalized trap strength $\Omega$ (c.f.~Eq.~(\ref{eq:Omega})).
Both characteristics are expected for the ground state within the polar phase
and thus corroborate the accuracy of our numerical BdG analysis. 

(ii) \emph{antiferromagnetic phase} --
The real and imaginary parts of all mode frequencies for the polar state in the antiferromagnetic phase, i.e., at parameters $(\delta,q)=(5 \cdot 10^{-3} , -0.1)$, are depicted in Fig.~\ref{fig:BdGPolarState}(a). 
We find that the polar state is dynamically unstable in this parameter regime as a continuum band of modes exhibits
$ |\Im (\omega) | >  10^{-4}$. 
The most unstable mode has a growth rate of $\gamma_{\mathrm{max}} = 0.1112$ which equals  $0.995 n_{\mathrm{p}}  \lvert \delta \rvert$.
Note that we use $n_{\mathrm{p}} \lvert \delta \rvert = 0.1118$ extracted by means of the ACTN method in \Sect{NumResGS} for comparison.
We find that this growth rate coincides with the growth rate for the homogeneous system resulting from Eq.~(\ref{eq:gammafPolar1}) when replacing the homogeneous density $n_{\mathrm{h}} $ by the peak density $n_{\mathrm{p}} $ of the trapped system.
This indicates that the peak density plays a crucial role in characterizing the stability properties of the trapped spin-1 system.
We will investigate this key observation in more detail below. 

The above stated property might suggest that the trap has no influence at all on the stability properties of the ground states. 
Nevertheless, we observe that the trap introduces new features in the system. 
To give an illustrative example we perform the BdG analysis of the polar state within the antiferromagnetic phase for parameters $-0.35 \leq q/ n_{\mathrm{p}} \delta \leq 0$.
For each value of $q$ we extract the squared mode frequencies
$\omega^2$ of the lowest $n$ eigenmodes of the trapped system. 
To compare the numerically obtained results with the homogeneous setting we calculate the squared mode frequencies
by means of \Eq{DispersionPolar} using momenta $k_n = \pi n / L_\mathrm{b}$, with $n \geq 1$, associated with the $n$-th eigenmode in a one-dimensional box of length $L_\mathrm{b} = 2 \, R_{\mathrm{TF}}$, where $R_{\mathrm{TF}}$ is the Thomas-Fermi radius of the corresponding trapped system.
\Fig{EnergySpectrumPolarInAF} shows the squared frequencies for both settings in units of $n_{\mathrm{p}} \delta$ as a function of the mode number $n$.
Except for the most unstable momentum mode, we observe strong deviations when the trap is present. 
Most strikingly the growth rates of unstable modes are smaller than in the homogeneous case. 
However, the crossing point to the stable regime is not altered by the trap.
We remark that our findings are in agreement with recent results obtained for the squared mode frequencies in a one-dimensional trapped spin-1 system using parameters $q/ (n_\mathrm{p} \delta) \in \{ -0.05, 0, 0.05 \}$ \cite{2018NJPh...20i5003V}.

As the goal of this work is to mainly investigate the overall structure of the stability of the spin-1 ground states we will in the following focus on 
discussing the maximal growth rates allowing to distinguish between stable and unstable regimes as well as to 
determine the dominant contribution to the growth of  mode occupations in case of an instability.

(iii) \emph{easy-plane phase} --
Fig.~\ref{fig:BdGPolarState}(d) shows the results of the BdG analysis for the polar state in the easy-plane phase, i.e., at parameters $(\delta,q)=(-5 \cdot 10^{-3} , 0.1)$. 
We observe once again a band of unstable modes fulfilling the criterion $|\Im (\omega) | >  10^{-4}$. 
In this case the most unstable mode has a growth rate of  $\gamma_{\mathrm{max}}  = 0.1118 =  n_{\mathrm{p}}  \lvert \delta \rvert $. 
The trapped system exhibits exactly the growth rate expected in a homogeneous system given by Eq.~(\ref{eq:gammafPolar2}) when replacing the homogeneous density with the peak density of the trapped system.

(iv) \emph{easy-axis phase} --
The real and imaginary parts of the mode frequencies for the polar state in the easy-axis phase, i.e.,~at parameters $(\delta,q)=(-5 \cdot 10^{-3} , -0.1)$, are presented in Fig.~\ref{fig:BdGPolarState}(c). Once again, in agreement with theory, we find
the polar state to be dynamically unstable due to a band of modes with  $|\Im (\omega) | >  10^{-4}$.
As in the case above the most unstable mode has a growth rate of  $\gamma_{\mathrm{max}}= n_{\mathrm{p}}  \lvert \delta \rvert $. 
Furthermore, in the easy-axis phase, as well,
the growth rate corresponds to the homogeneous case (see  Eq.~(\ref{eq:gammafPolar2})) with the peak density replacing the homogeneous density.

Hence, we find that the stability properties of the most unstable mode, characterized by the maximal growth rate $\gamma_{\mathrm{max}}$, coincide with the homogeneous prediction in the above shown example. 
To show that this property is indeed valid over variations of parameters,
we carry out the BdG analysis of the polar state at various quadratic Zeeman energies. We
perform the relevant continuations
for fixed spin coupling $\delta$, but for both types of spin interactions. 

The maximal growth rate $\gamma_{\mathrm{max}} = \mathrm{max} \,  |\Im (\omega)|$ as a function of $q$ and different signs of $\delta$ is shown in Fig.~\ref{fig:MaxGrowthRatesPolarState}. 
To allow for a direct comparison to the homogeneous predictions the growth rates and the quadratic Zeeman energies are given in units of $n_{\mathrm{p}} \lvert \delta \rvert$.
For $\delta > 0$ the growth rate follows the homogeneous prediction for the whole parameter range.
The same feature is found in case of $\delta < 0$. 
The observed behavior for $\delta > 0$ coincides with the one for $\delta <0$ when shifting the quadratic Zeeman energy by two units. This is in exact agreement with the shift of the phase transition from $q/(n_{\mathrm{p}} \lvert \delta \rvert) =0$ to $q/(n_{\mathrm{p}} \lvert \delta \rvert) = 2$ which shows that the exact same properties are found irrespective of the sign of $\delta$.

\begin{figure}
\includegraphics{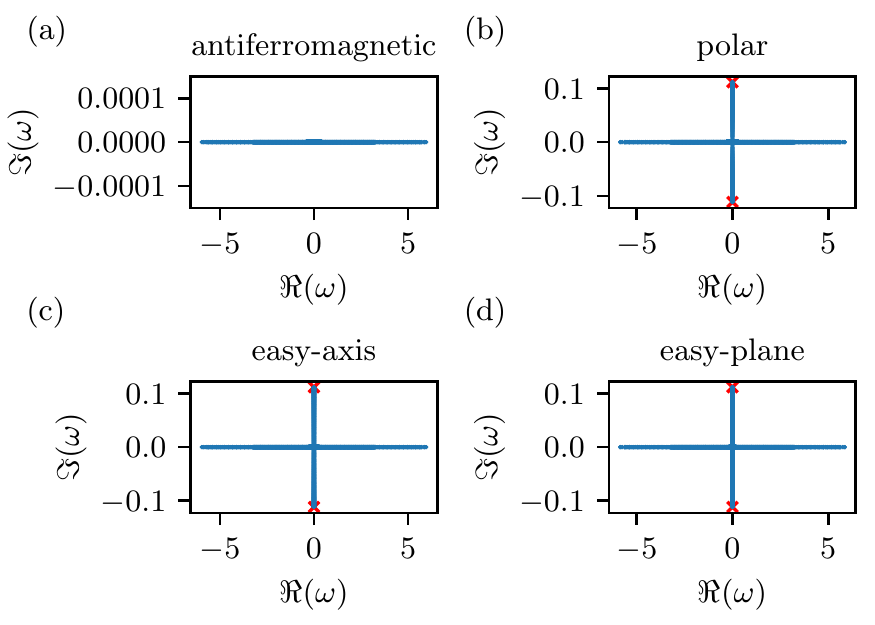}
\caption{\label{fig:BdGAFState}  Real ($\Re$) and imaginary ($\Im$) parts of the mode frequencies $\omega$ resulting from the BdG analysis of the \emph{antiferromagnetic} state $\psi \sim \left(1,0,1 \right)^T$ within the (a) antiferromagnetic $(\delta, q) = ( 5\cdot 10^{-3} , -0.1)$, (b) polar $( 5\cdot 10^{-3} , 0.1)$, (c) easy-axis $( -5\cdot 10^{-3} , -0.1)$ and (d) easy-plane $(- 5\cdot 10^{-3} , 0.1)$ phase. 
In panels (b), (c) and (d) we see a continuum band of mode frequencies along the real and the imaginary axis indicating that the frequencies are either purely real or purely imaginary.
The red crosses mark the predicted mode frequency with the largest imaginary part within the given parameter regime derived from homogeneous Bogoliubov theory  by replacing the homogeneous density by the peak density of the trapped system. 
The prediction is in good agreement with the largest imaginary part of the numerically obtained mode frequencies.}
\end{figure}

\begin{figure}
\includegraphics{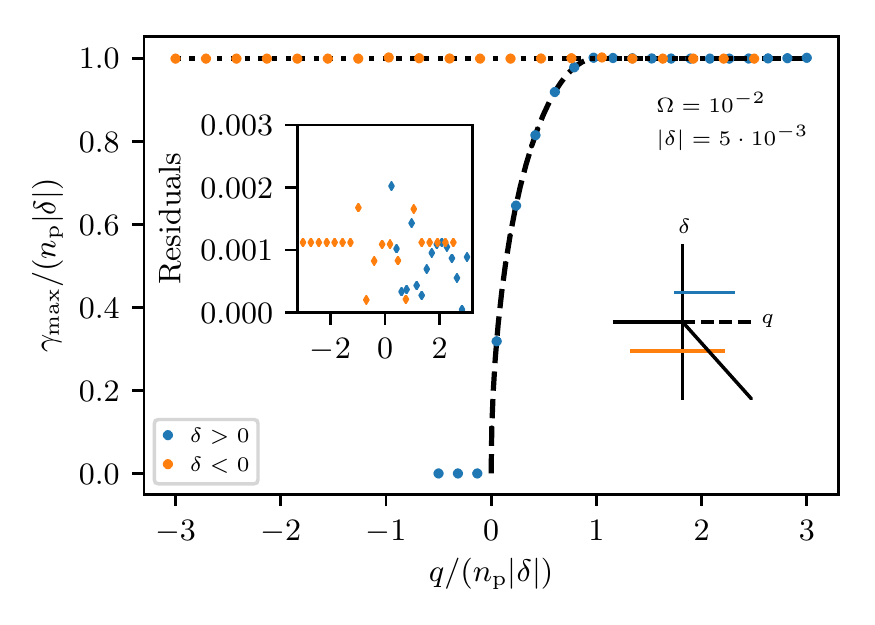}
\caption{\label{fig:MaxGrowthRatesAFState} Maximal growth rate $\gamma_{\mathrm{max}} = \mathrm{max}\, | \Im (\omega) |$ of the eigenmodes obtained by means of the BdG analysis of the \emph{antiferromagnetic} state as a function of $q$ and the sign of $\delta$.  
The investigated parameter regime is indicated by the blue and orange solid lines in the schematic representation of the spin-1 phase diagram (c.f.~\Fig{PhaseDiag}).  
The analysis is performed for parameters $\Omega = 10^{-2}$ and $\lvert \delta \rvert = 5  \cdot 10^{-3}$. 
The growth rate as well as the quadratic Zeeman energy are given in units of $n_{\mathrm{p}} \lvert \delta \rvert$. 
For $\delta > 0$ the growth rate follows the homogeneous prediction (dashed line) for the whole parameter range. 
In case of $\delta < 0$ we observe a constant growth rate of $1$ irrespective of $q$. 
The growth rate coincides with the homogeneous setting (dotted line). The antiferromagnetic state is always dynamically unstable for $\delta <0$.
The inset shows the residuals $|\gamma_{\mathrm{max, h}} - \gamma_{\mathrm{max}}|/ \gamma_{\mathrm{max,h}}$, with $\gamma_{\mathrm{max,h}}$ being the growth rate calculated from the homogeneous prediction. 
The deviation of the numerically extracted growth rates from the homogeneous prediction is less than $0.2 \%$ for all parameters considered. The color coding is as in the main frame.} 
\end{figure}

\subsubsection{Excitations about the antiferromagnetic state}
\label{sec:AFState}

We continue by investigating the dynamical stability of the antiferromagnetic state $\psi \sim \left(1,0,1 \right)^T$ throughout the phases of our trapped spin-1 system. 
The initial state for the wave functions used in the Newton scheme is taken to be a Gaussian, centered around the middle of the trap, with width $\sigma = 500 / \sqrt{2}$ in the $m_{\mathrm{F}} = \pm 1$ components and $0$ in the $m_{\mathrm{F}} = 0$ component. 
The Newton method converges to the antiferromagnetic state in all phases within 9 iterations. 
The error tolerance is set to $10^{-10}$ as before.

(i) \emph{antiferromagnetic phase} --
Fig.~\ref{fig:BdGAFState}(a) shows the real and imaginary parts of the mode
frequencies obtained by means of the BdG analysis in the antiferromagnetic phase.
We find no dynamically unstable modes.
As before, this is the consistency check of the method as the antiferromagnetic state is the ground state in this phase and thus has to be stable.

(ii) \emph{polar phase} --
Within the polar phase (see real and imaginary parts of the mode frequencies
depicted in Fig.~\ref{fig:BdGAFState}(b)) the antiferromagnetic state is dynamically unstable as we observe
a band of eigenfrequencies with $ |\Im (\omega) | >  10^{-4}$. 
The growth rate of the most unstable mode is  $\gamma_{\mathrm{max}} = 0.995 n_{\mathrm{p}} \lvert \delta \rvert$, once again
coinciding with the growth rate for the homogeneous system
obtained through Eq.~(\ref{eq:gamma0AF1}) by replacing the homogeneous density $n_{\mathrm{h}} $ with the peak density $n_{\mathrm{p}} $ of  the trapped system. 

(iii) \emph{easy-axis phase} --
Fig.~\ref{fig:BdGAFState}(c) shows the real and imaginary parts of the mode
frequencies resulting from the BdG analysis of the antiferromagnetic state in the easy-axis phase, where again a band of unstable eigenmodes arises. 
The most unstable one exhibits a growth rate of $\gamma_{\mathrm{max}} = n_{\mathrm{p}} \lvert \delta \rvert$.  
In this parameter regime the growth rate equals the homogeneous prediction given in Eq.~(\ref{eq:gammamagAF}) when exchanging the homogeneous density with the
trapped problem peak density.

(iv) \emph{easy-plane phase} --
Last, we present the real and imaginary parts of the mode frequencies
for the antiferromagnetic state in the easy-plane phase in Fig.~\ref{fig:BdGAFState}(d). 
The instability here is found to possess a maximal
growth rate of $\gamma_{\mathrm{max}} =  n_{\mathrm{p}} \lvert \delta \rvert$, in line with Eq.~(\ref{eq:gammamagAF}). 

To look for further agreement with the homogeneous setting we investigate the stability properties of the antiferromagnetic state for various quadratic Zeeman energies and different signs of $\delta$.
The maximal growth rate $\gamma_{\mathrm{max}} = \mathrm{max} \, | \Im (\omega) |$ as a function of $q$ and both signs of $\delta$ is shown in Fig.~\ref{fig:MaxGrowthRatesAFState}. 
To allow for a direct comparison to the homogeneous predictions the growth rates and the quadratic Zeeman energies are given in units of $n_{\mathrm{p}} \lvert \delta \rvert$.
For $\delta > 0$ the growth rate follows the homogeneous prediction for the whole parameter range.
In case of $\delta < 0$ we observe a constant growth rate of $1$ throughout the whole parameter range.
This coincides with the homogeneous setting where the maximal growth rate is always $1$ in the units used here for $\delta < 0$ (c.f.~Eqs.~(\ref{eq:gammamagAF}) and (\ref{eq:gamma0AF2})).
Our BdG analysis shows furthermore that the antiferromagnetic state is always dynamically unstable for $\delta <0$.

\begin{figure}
\includegraphics{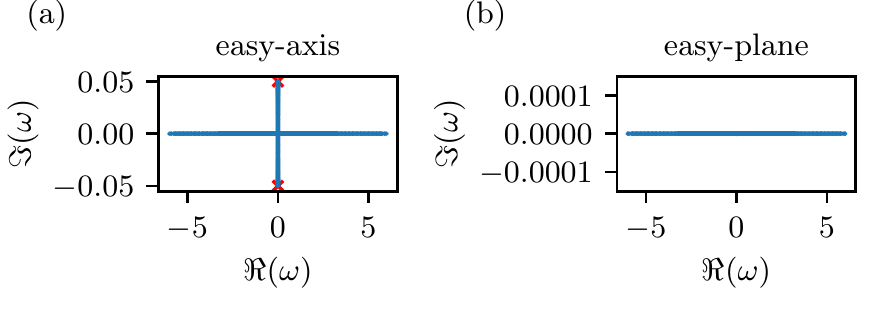}
\caption{\label{fig:BdGEPState}  Real ($\Re$) and imaginary ($\Im$) parts of the mode frequencies $\omega$ resulting from the BdG analysis of the \emph{easy-plane} state $\psi \sim \left (\sqrt{1-q/q_0}, \sqrt{2(1 +q/q_0)}, \sqrt{1 -q/q_0} \right)^T$ within the (a) easy-axis $(\delta, q ) = ( -5\cdot 10^{-3} , -0.1)$ and (b) easy-plane $(- 5\cdot 10^{-3} , 0.1)$ phase. 
In panel (a) we see a continuum band of mode frequencies along the real and the imaginary axis indicating that the frequencies are either purely real or purely imaginary.
The red crosses mark the predicted mode frequency with the largest imaginary part within the given parameter regime derived from homogeneous Bogoliubov theory.
The numerically obtained largest imaginary part of the mode energies is in good agreement with the homogeneous prediction given by $\lvert q \rvert /2 = 0.05$.}
\end{figure}

\begin{figure}
\includegraphics{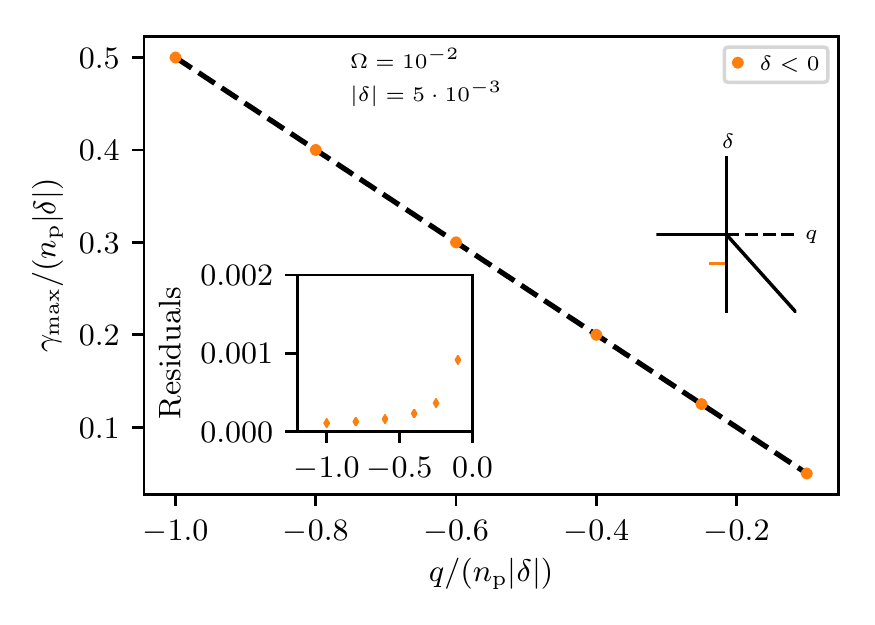}
\caption{\label{fig:MaxGrowthRatesEPState} Maximal growth rate $\gamma_{\mathrm{max}} = \mathrm{max} \,  | \Im (\omega ) |$ of the eigenmodes obtained by means of the BdG analysis of the \emph{easy-plane} state as a function of $q$ and $\delta < 0$.  
The investigated parameter regime is indicated by the orange solid line in the schematic representation of the spin-1 phase diagram (c.f.~\Fig{PhaseDiag}).
The analysis is performed for parameters $\Omega = 10^{-2}$ and $\lvert \delta \rvert = 5  \cdot 10^{-3}$. 
The growth rate as well as the quadratic Zeeman energy are given in units of $n_{\mathrm{p}} \lvert \delta \rvert$. 
The growth rate follows the homogeneous prediction (dashed line) given by $\gamma_{\mathrm{max}} = \lvert q \rvert /2$ over the whole parameter range.
The inset shows the residuals $|\gamma_{\mathrm{max, h}} - \gamma_{\mathrm{max}}|/ \gamma_{\mathrm{max,h}}$, with $\gamma_{\mathrm{max,h}}$ being the growth rate calculated from the homogeneous prediction. 
The deviation of the numerically extracted growth rates from the homogeneous prediction is less than $0.1 \%$ for all parameters considered. The color coding is as in the main frame.} 
\end{figure}

\subsubsection{Excitations about the easy-plane state}
\label{sec:EPState}

We proceed by investigating the dynamical stability of the easy-plane state $\psi \sim \left (\sqrt{1-q/q_0}, \sqrt{2(1 +q/q_0)}, \sqrt{1 -q/q_0} \right)^T$  within the easy-plane and the easy-axis phase of our trapped spin-1 system. 
The BdG analysis for $\delta >0$ is not shown here as we were not able to converge to the easy-plane state in the antiferromagnetic and polar phase by means of a standard Newton method.
The initial state  for the wave functions used in the Newton scheme within the easy-axis phase is taken to be a Gaussian,  centered around the middle of the trap, with width $\sigma = 500 / \sqrt{2}$  where the amplitude in the $m_{\mathrm{F}} = 0$ component is a factor of $1/\sqrt{2}$ smaller than in the $m_{\mathrm{F}} = \pm 1$ components.
Within the easy-plane phase the initial Gaussian also has a width of $\sigma = 500 / \sqrt{2}$, however the amplitudes  in the $m_{\mathrm{F}} = \pm 1$ components are a factor of two smaller than in the $m_{\mathrm{F}} = 0$ component.
The Newton method converges to the easy-plane state in both phases within 18 iterations.
The error tolerance is again set to $10^{-10}$. It is relevant to mention
here that this easy-plane state is the {\it only} one for which
we are not able to converge to it, throughout the parametric variations
that we considered (due to the absence of convergence in the $\delta>0$
regime).

(i) \emph{easy-plane phase} --
The real and imaginary parts of the mode frequencies of the easy-plane state in the easy-plane phase are depicted in Fig.~\ref{fig:BdGEPState}(b).
The state is dynamically stable in this parameter regime, as expected
from its ground state nature in this regime.

(ii) \emph{easy-axis phase} --
Fig.~\ref{fig:BdGEPState}(a) shows the real and imaginary parts of the mode frequencies resulting from the BdG analysis of the easy-plane state in the easy-axis phase. 
In this case, a band of dynamically unstable modes  with $|\Im (\omega) | >  10^{-4}$ arises.
The growth rate of the most unstable mode is given by $\gamma_{\mathrm{max}} = 0.05 = \lvert q\rvert /2$. 
This coincides with the prediction for the homogeneous system stated in Eq.~(\ref{eq:gamm0EP}). 

To confirm that the maximal growth rate indeed follows a linear function in $\lvert  q \rvert$ we perform the BdG analysis for various negative quadratic Zeeman energies.  
We present the results obtained for $- n_{\mathrm{p}} \lvert \delta \rvert  \leq q \leq -0.1 n_{\mathrm{p}} \lvert \delta \rvert$. 
The initial guess for the Newton method has to be adjusted to reflect the final population of the three components.
The maximal growth rate $\gamma_{\mathrm{max}} = \mathrm{max} \, | \Im (\omega ) |$ agrees exactly with  $\lvert q\rvert /2$ (see dashed line in Fig.~\ref{fig:MaxGrowthRatesEPState}). The growth rates and the quadratic Zeeman energies are again given in units of $n_{\mathrm{p}} \lvert \delta \rvert$.

\subsubsection{Excitations about the easy-axis state}
\label{sec:EAState}

We finally study the stability properties of the easy-axis state $\psi \sim \left(1,0,0 \right)^T$ throughout the different phases of our trapped spin-1 system. 
The initial state of the wave function of the $m_{\mathrm{F}} = 1$ component used in the Newton scheme is taken to be a Gaussian, centered around the middle of the trap, with width $\sigma = 500 / \sqrt{2}$. 
In addition we start with a vanishing wave function in the $m_{\mathrm{F}} = 0, -1$ components. 
The Newton method converges to the easy-axis state in all phases within 9 iterations. 
The error tolerance is again set to $10^{-10}$.

\begin{figure}
\includegraphics{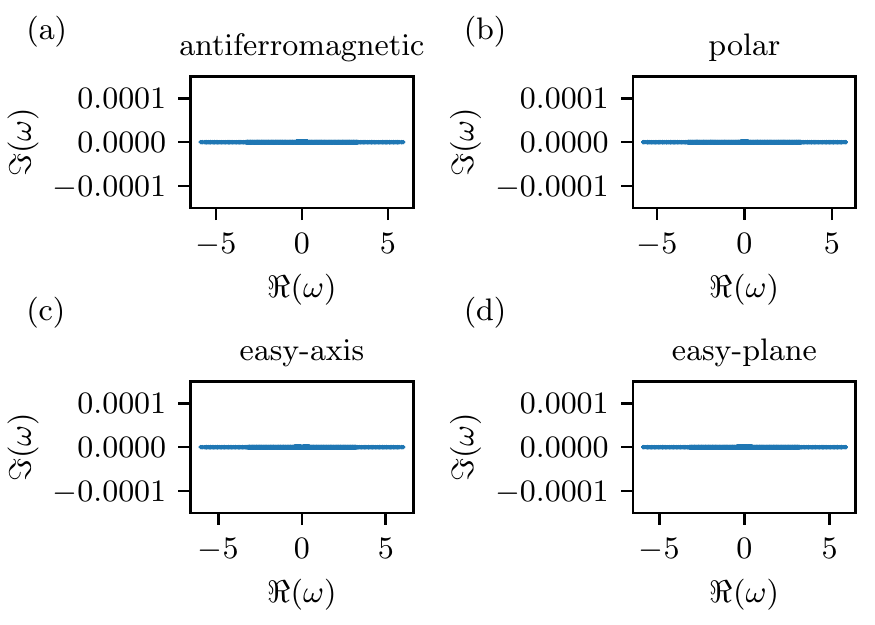}
\caption{\label{fig:BdGEAState} Real ($\Re$) and imaginary ($\Im$) parts of the mode frequencies $\omega$ resulting from the BdG analysis of the \emph{easy-axis} state $\psi \sim \left(1,0,0 \right)^T$ within the (a) antiferromagnetic $(\delta, q) = ( 5\cdot 10^{-3} , -0.1)$, (b) polar $( 5\cdot 10^{-3} , 0.1)$, (c) easy-axis $( -5\cdot 10^{-3} , -0.1)$ and (d) easy-plane $(- 5\cdot 10^{-3} , 0.1)$ phase. The easy-axis state is stable in all phases. This result agrees with the prediction for the homogeneous system in case of $\lvert \delta \vert =  5 \cdot 10^{-3}$.}
\end{figure}

Fig.~\ref{fig:BdGEAState} shows the real and the imaginary parts of the mode frequencies obtained by means of the BdG analysis of the easy-axis state in the different phases. 
We find that the easy-axis state is stable in all phases. 
This agrees with the prediction for the homogeneous system in the case of $\lvert \delta \vert =  5 \cdot 10^{-3}$.

\subsubsection{Role of spin coupling strength $\delta$}
\label{sec:RoleOfDelta}

As we observe the stability properties of the ground states in the trapped system to match the homogeneous setting we wish to investigate the dependence of those properties on the strength of the spin coupling $\delta$.
This is of particular interest as different magnitudes of the spin coupling are realized in experiments.
We therefore increase the spin coupling by a factor of 4  such that $\lvert \delta \rvert = 2 \cdot 10^{-2}$ which is close to the experimental coupling for sodium.
To resolve all unstable momentum modes on our numerical grid properly we have to increase the number of grid points to $N_\mathrm{g }= 1024$. 
All other parameters remain unchanged. 
As a prototypical example, we investigate the maximal growth rates obtained by means of the BdG analysis of the polar state as a function of the quadratic Zeeman energy $q$ and different signs of $\delta$.

\begin{figure}
\includegraphics{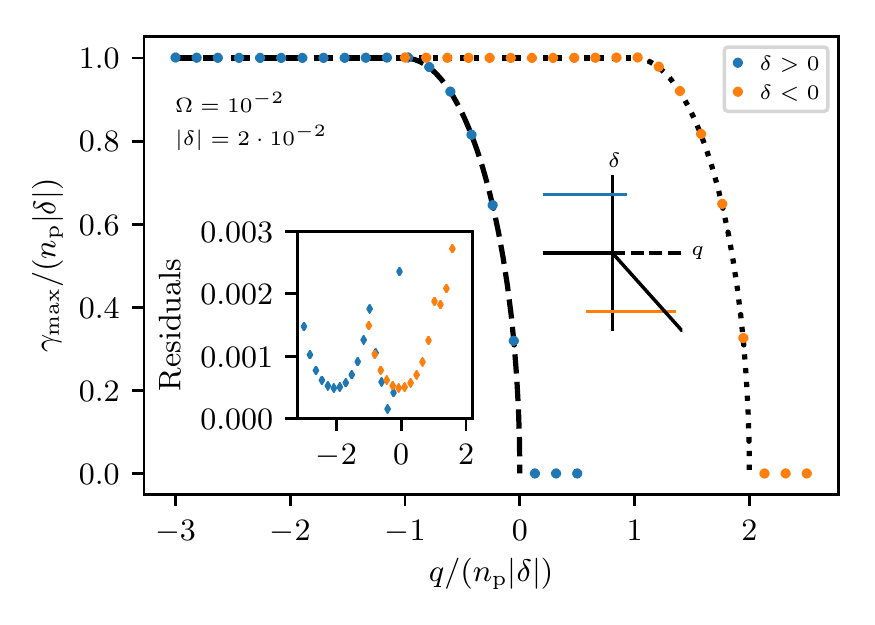}
\caption{\label{fig:MaxGrowthRatesPolarStateDeltaFactor4Larger} Maximal growth rate $\gamma_{\mathrm{max}} = \mathrm{max} \, | \Im (\omega ) |$ of the eigenmodes obtained by means of the BdG analysis of the \emph{polar} state  as a function of $q$ and the sign of $\delta$.  
The investigated parameter regime is indicated by the blue and orange solid lines in the schematic representation of the spin-1 phase diagram (c.f.~\Fig{PhaseDiag}).  
The analysis is performed for parameters $\Omega = 10^{-2}$ and $\lvert \delta \rvert = 2  \cdot 10^{-2}$, i.e., the strength of the spin coupling is increased by a factor of 4 as compared to \Fig{MaxGrowthRatesPolarState}. 
The growth rate as well as the quadratic Zeeman energy are given in units of $n_{\mathrm{p}} \lvert  \delta \rvert$.  
For $\delta > 0$ the growth rate follows the homogeneous prediction (dashed line) for the whole parameter range. 
The same characteristics appear in case of $\delta < 0$ where the data is matched by the dotted lines showing the homogeneous case.
The inset shows the residuals $|\gamma_{\mathrm{max, h}} - \gamma_{\mathrm{max}}|/ \gamma_{\mathrm{max,h}}$, with $\gamma_{\mathrm{max,h}}$ being the growth rate calculated from the homogeneous prediction. 
The deviation of the numerically extracted growth rates from the homogeneous prediction is less than $0.3 \%$ for all parameters considered. The color coding is as in the main frame.} 
\end{figure}

Fig.~\ref{fig:MaxGrowthRatesPolarStateDeltaFactor4Larger} shows the maximal growth rate $\gamma_{\mathrm{max}} = \mathrm{max}  \, | \Im (\omega) |$  for $\lvert \delta \rvert = 2 \cdot 10^{-2}$. 
The growth rate as well as the quadratic Zeeman energy are given in units of $n_{\mathrm{p}} \lvert  \delta \rvert$.  
The growth rate follows the homogeneous prediction for the whole parameter range, both for the positive and
for the negative value of $\delta$ (see dashed and dotted line respectively).

Our analysis indicates that the stability properties  of the ground states do not change when increasing the spin coupling strength. 
Thus we expect to observe the same dynamical instabilities in systems with different spin couplings when choosing the quadratic Zeeman energy $q$ in the corresponding units of $n_{\mathrm{p}}  \lvert \delta \rvert$.

\begin{figure}
\includegraphics{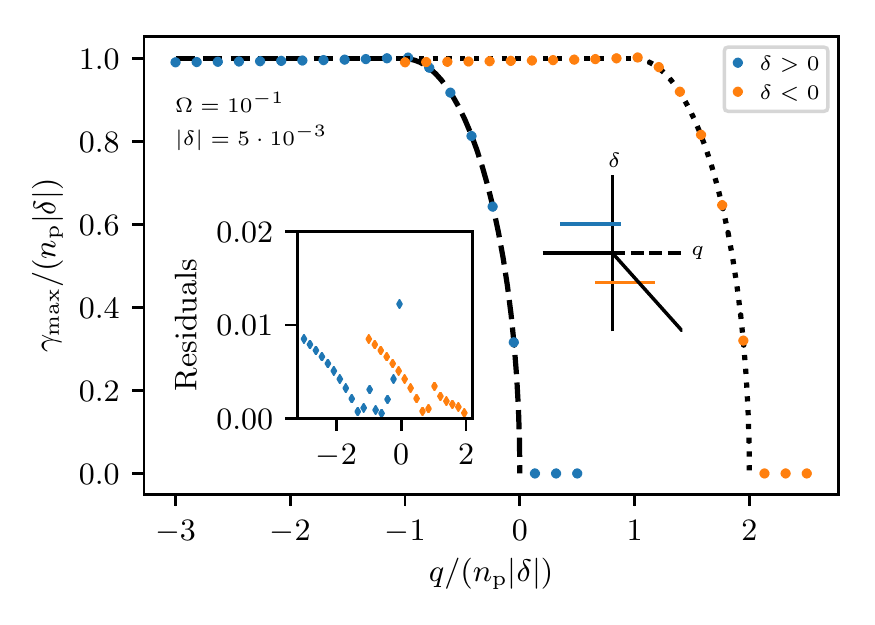}
\caption{\label{fig:MaxGrowthRatesPolarStateOmegaFactor10Larger} Maximal growth rate $\gamma_{\mathrm{max}} = \mathrm{max} \,  | \Im (\omega) |$ of the eigenmodes obtained by means of the BdG analysis of the \emph{polar} state  as a function of $q$ and the sign of $\delta$.  
The investigated parameter regime is indicated by the blue and orange solid lines in the schematic representation of the spin-1 phase diagram (c.f.~\Fig{PhaseDiag}). 
The analysis is performed for parameters $\Omega = 10^{-1}$ and $\lvert \delta \rvert = 5  \cdot 10^{-3}$ ,i.e., the normalized trap strength is increased by a factor of 10 as compared to \Fig{MaxGrowthRatesPolarState}. 
The growth rate as well as the quadratic Zeeman energy are given in units of $n_{\mathrm{p}} \lvert  \delta \rvert$.  For $\delta > 0$ the growth rate follows the homogeneous prediction (dashed line) for $q>-1$. 
For $q<-1$  the growth rate tends to attain smaller values than in the homogeneous case (dashed line) showing a maximal deviation of about $1 \%$. 
The same phenomenon appears in case of $\delta < 0$ where the data is matched by the dotted lines for $q>1$. 
The growth rate again shows a maximal deviation of about $1 \%$ from the homogeneous case (dotted line) for $q<1$. 
The deviations are more clearly visible in the inset which shows the residuals $|\gamma_{\mathrm{max, h}} - \gamma_{\mathrm{max}}|/ \gamma_{\mathrm{max,h}}$, with $\gamma_{\mathrm{max,h}}$ being the growth rate calculated from the homogeneous prediction. The color coding is as in the main frame.
Increasing the normalized trap strength by an order of magnitude only causes minor changes of the stability properties. This is however expected as we are about to leave the one-dimensional regime as the transversal trapping frequency becomes comparable to the longitudinal trapping frequency.} 
\end{figure}

\begin{figure}
\includegraphics{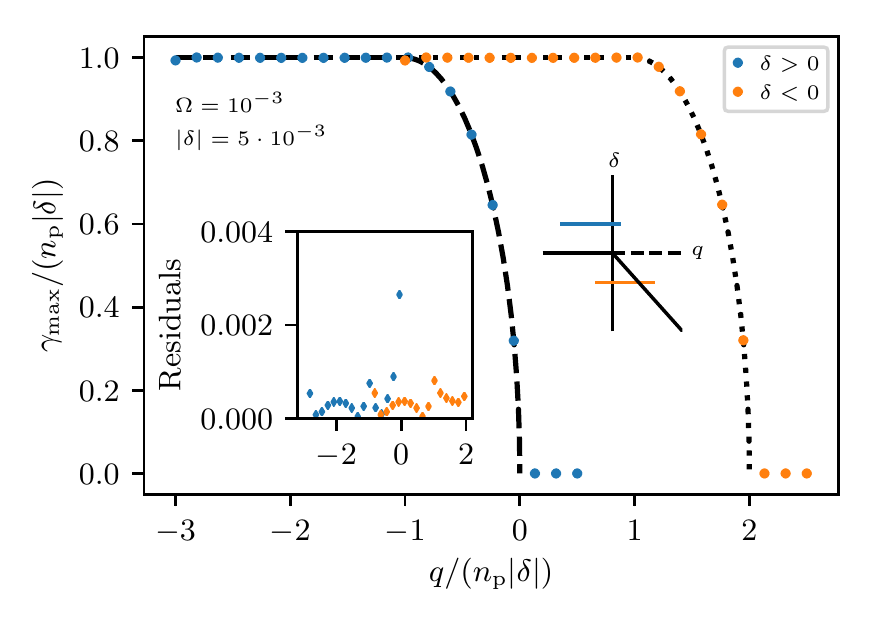}
\caption{\label{fig:MaxGrowthRatesPolarStateOmegaFactor10Smaller} Maximal growth rate $\gamma_{\mathrm{max}} = \mathrm{max} \, | \Im (\omega) |$ of the eigenmodes obtained by means of the BdG analysis of the \emph{polar} state  as a function of $q$ and the sign of $\delta$.  
The investigated parameter regime is indicated by the blue and orange solid lines in the schematic representation of the spin-1 phase diagram (c.f.~\Fig{PhaseDiag}).  
The analysis is performed for parameters $\Omega = 10^{-3}$ and $\lvert \delta \rvert = 5  \cdot 10^{-3}$ ,i.e., the normalized trap strength is decreased by a factor of 10 as compared to \Fig{MaxGrowthRatesPolarState}. 
The growth rate as well as the quadratic Zeeman energy are given in units of $n_{\mathrm{p}} \lvert  \delta \rvert$.  For $\delta > 0$ the growth rate follows the homogeneous prediction (dashed line) for the whole parameter range. 
The same phenomenon appears in case of $\delta < 0$ where the data is matched by the dotted line. 
Decreasing the normalized trap strength by an order of magnitude does not change the stability properties (c.f.~\Fig{MaxGrowthRatesPolarState}). 
The observed behavior is expected as we are further approaching the homogeneous setting when lowering $\Omega$.
The inset shows the residuals $|\gamma_{\mathrm{max, h}} - \gamma_{\mathrm{max}}|/ \gamma_{\mathrm{max,h}}$, with $\gamma_{\mathrm{max,h}}$ being the growth rate calculated from the homogeneous prediction. 
The deviation of the numerically extracted growth rates from the homogeneous prediction is less than $0.3 \%$ for all parameters considered. Color coding as in the main frame.} 
\end{figure}

\subsubsection{Role of normalized trap strength $\Omega$}
\label{sec:RoleOfOmega}

In the following we want to study the role of the normalized trap strength $\Omega$ on the stability properties.
All results discussed in this part are obtained for a spin coupling of $ \lvert \delta \rvert = 5 \cdot 10^{-3}$.
The homogeneous setting is recovered as $\Omega \rightarrow 0$. 
On the other hand, when $\Omega$ is increased,
we progressively depart from  the 1D regime as the condition $\omega_{\perp} \gg \omega_{\parallel}$ is not fulfilled anymore.
When the transverse and longitudinal trapping frequencies become comparable in magnitude we expect deviations of the stability properties from the previously shown ones as we have to include additional transversal degrees of freedom. 
We therefore choose the parameter $\Omega$ to be sufficiently far away from unity, so as to remain within a quasi-1D realm. 
Here, too, we use $N_{\mathrm{g}}=1024$.

We start by increasing $\Omega$ by a factor of 10 such that $\Omega =10^{-1}$. 
Fig.~\ref{fig:MaxGrowthRatesPolarStateOmegaFactor10Larger} shows the maximal growth rate $\gamma_{\mathrm{max}} = \mathrm{max}  \,| \Im (\omega )|$ of the eigenmodes as a function of $q$ and the sign of $\delta$.
The growth rate as well as the quadratic Zeeman energy are given in units of $n_{\mathrm{p}} \lvert  \delta \rvert$.
For $\delta > 0$ the growth rate follows the homogeneous prediction (dashed line) for $q > -1$. For $q < -1$ the growth rate is slightly smaller than in the homogeneous case (dashed line) with a maximal deviation of about $1 \%$.
The same feature appears in case of $\delta < 0$. 
Here the growth rate agrees with the homogeneous prediction (dotted line) for $q > 1$. The rate also shows a maximal deviation of about $1 \%$ from the
homogeneous case (dotted line) for $q < 1$.

We continue with analyzing the stability properties of the polar state when decreasing the normalized trap strength by a factor of 10 such that $\Omega = 10^{-3}$.
The maximal growth rate $\gamma_{\mathrm{max}} = \mathrm{max}  \, | \Im (\omega) |$ of the eigenmodes obtained by means of the BdG analysis as a function of $q$ and the sign of $\delta$  is depicted in Fig.~\ref{fig:MaxGrowthRatesPolarStateOmegaFactor10Smaller}. 
The growth rate as well as the quadratic Zeeman energy are given in units of $n_{\mathrm{p}} \lvert  \delta \rvert$.  
For $\delta > 0$ the growth rate follows the homogeneous prediction (dashed line) for the whole parameter range. 
The same characteristics appear in case of $\delta < 0$ (see dotted line).

We conclude that increasing the normalized trap strength by an order of magnitude leads to small deviations from the homogeneous predictions when $q<-1$ for $\delta>0$ and $q< 1$ 
for $\delta<0$.
Beyond this trapping strength, one progressively departs from the one-dimensional regime as the transversal trapping frequency becomes comparable to the longitudinal one.
We find no deviations of the stability properties from the previously discussed setting where $\Omega = 10^{-2}$ when decreasing the normalized trap strength by an order of magnitude.  
This is expected as we are approaching the homogeneous setting even further by lowering the normalized trap strength. 

Our analysis shows that the principal stability properties of the ground states agree with the homogeneous setting over several orders of magnitude of the normalized trap strength $\Omega$.
Thus we expect the characteristics of dynamical instabilities to agree in various one-dimensional trapping geometries.  
The trapping potential is found to not alter the overall stability properties calculated for homogeneous settings.
It solely introduces the peak density of the trapped system
which appears
in the equations for the maximal growth rates instead of the homogeneous density.
Deviations will, naturally, appear progressively as the trapping strength
increases towards values closer to $\Omega \rightarrow 1$.

\section{Conclusion and Outlook} \label{sec:Conclusion}

In this work we studied the stability properties of the ground states of a trapped one-dimensional spin-1 Bose gas. 
We started by mapping out the ground state phase diagram of the trapped system.
Therefore we made use of the accelerated continuous-time Nesterov
(ACTN) method which we extended to our multi-component system.
We showed that the ACTN method is a robust and powerful tool for finding the ground states of a physical system as it does not require a highly
accurate initial guess for the wave function of the different components.
This makes the method extremely useful to explore systems with unknown phase diagrams in the future.

We numerically performed a stability analysis of the spin-1 ground states by solving the BdG equations for the trapped system.
We found that the principal stability conclusions  for
the ground states coincide with
the predictions made in absence of a trapping potential, although
as shown in \Fig{EnergySpectrumPolarInAF}, the spectrum is not identical
and the growth rates of modes other than the most unstable one are indeed altered. 
The maximal growth rates  obtained in the trapped system match the homogeneous predictions when replacing the homogeneous density with the peak density of the trapped system in the corresponding equations. 
The near-independence of the stability conclusions is valid within the regime of
quasi-1D values of the normalized trap strength (representing the ratio of longitudinal
to transverse trapping frequencies). It should be noted that
we explored each of the possible states (polar, antiferromagnetic,
easy-plane and easy-axis) in almost each of the possible regimes,
identifying the states in the regimes where they are no longer
the ground state via Newton iterations. In the latter cases,
potential instabilities of the states were elucidated.

Naturally, this work paves the way for numerous additional investigations
of interest for the near future. For instance, the question was raised
through our studies of whether the easy-plane state can be found
to exist in the half-plane with $\delta>0$. Moreover, for the 1D setting,
we tackled the ground states of the system and their stability over
the $(q,\delta)$ space. Nevertheless, there are numerous intriguing excited
states, including ones involving solitary waves that are experimentally
accessible~\cite{PhysRevLett.120.063202}.
It would be particularly relevant to extend our techniques to
the latter context.
Finally, while here we focused on the quasi-1D setting,
naturally, adapting such techniques to 2D and 3D spinorial
states would be of interest in its own right. Some of these
directions are currently under examination and will be reported
on in future publications.

\begin{acknowledgments}
The authors thank C.~B.~Ward, M.~Pr\"ufer and the SynQS-Team in Heidelberg for discussions and collaboration on the topics described here. 
This work was supported by the Heidelberg Graduate School of Fundamental Physics (HGSFP), the Heidelberg Center for Quantum Dynamics (CQD), the European Commission, within the Horizon-2020 programme, through the FET-Proactive grant AQuS (Project No. 640800) and the ERC Advanced Grant EntangleGen (Project-ID 694561) as well as the DFG Collaborative Research Center SFB1225 (ISOQUANT).
C.-M.S. thanks the Department of Mathematics and Statistics, University of Massachusetts, USA, for hospitality and support.
This material is based upon work supported by the U.S.\ National Science Foundation under Grant No.\ PHY-1602994 and by the Alexander von Humboldt Foundation (P.G.K.).
\end{acknowledgments}

\bibliographystyle{apsrev4-1}

%



\end{document}